\documentclass[10pt,conference]{IEEEtran}
\IEEEoverridecommandlockouts
\usepackage{balance}
\usepackage{float}
\usepackage{cite}
\usepackage{amsmath,amssymb,amsfonts}
\usepackage{textcomp}
\usepackage{xcolor}
\usepackage{colortbl}
\usepackage{booktabs}
\usepackage{color}
\usepackage{amsmath}
\usepackage[linesnumbered,ruled,vlined]{algorithm2e}
\usepackage{hyperref}
\usepackage{cleveref}
\hypersetup{hidelinks}
\usepackage{multirow}
\usepackage{booktabs}
\usepackage{bm}
\usepackage[normalem]{ulem}

\usepackage{tikz}
\usepackage{pgfplots}
\usepackage{microtype}
\usepackage{graphicx}
\usepackage{adjustbox}
\usepackage{xspace}
\usepackage{subcaption}
\usepackage[tablename=Table]{caption}
\usepackage{thmtools, thm-restate}
\usepackage{pifont}
\usepackage{listings}
\usepackage{threeparttable}
\usepackage{colortbl}

\ifCLASSOPTIONcompsoc

\else

\fi

\def\BibTeX{{\rm B\kern-.05em{\sc i\kern-.025em b}\kern-.08em
    T\kern-.1667em\lower.7ex\hbox{E}\kern-.125emX}}

\begin{document}
\title{FaultGPT: Industrial Fault Diagnosis Question Answering System by Vision Language Models}
\author{
\IEEEauthorblockN{
Jiao Chen,
Ruyi Huang,
Zuohong Lv,\\
Jianhua Tang,~\IEEEmembership{Senior~Member,~IEEE}, 
and Weihua Li~\IEEEmembership{Senior~Member,~IEEE}
}
\thanks{This research work has been submitted to IEEE for peer review and potential publication. Should the copyright be transferred, this version may become inaccessible without prior notice.}
\thanks{
Jiao Chen, Ruyi Huang, Zuohong Lv, and Jianhua Tang are with the Shien-Ming Wu School of Intelligent Engineering, South China University of Technology, Guangzhou 511442, China. 
Weihua Li is with the School of Mechanical and Automotive Engineering, South China University of Technology, Guangzhou 510641, China. 
Jianhua Tang and Weihua Li are also with Pazhou Lab, Guangzhou 510335, China.
Email addresses: 202110190459@mail.scut.edu.cn, snowxiaoyu@hotmail.com, 202220159664@mail.scut.edu.cn, jtang4@e.ntu.edu.sg, whlee@scut.edu.cn.
The corresponding author is Jianhua Tang. 
}
\thanks{The project page: \url{https://sites.google.com/view/faultgpt}.}
}
\maketitle

\begin{abstract}
Recently, employing single-modality large language models based on mechanical vibration signals as \textit{Tuning Predictors} has introduced new perspectives in intelligent fault diagnosis.
However, the potential of these methods to leverage multimodal data remains underexploited, particularly in complex mechanical systems where relying on a single data source often fails to capture comprehensive fault information.
In this paper, we present FaultGPT, a novel model that generates fault diagnosis reports directly from raw vibration signals. By leveraging large vision-language models (LVLM) and text-based supervision, FaultGPT performs end-to-end fault diagnosis question answering (FDQA), distinguishing itself from traditional classification or regression approaches.
Specifically, we construct a large-scale FDQA instruction dataset for instruction tuning of LVLM. This dataset includes vibration time-frequency image-text label pairs and human instruction-ground truth pairs.
To enhance the capability in generating high-quality fault diagnosis reports, we design a multi-scale cross-modal image decoder to extract fine-grained fault semantics and conducted instruction tuning without introducing additional training parameters into the LVLM.
Extensive experiments, including fault diagnosis report generation, few-shot and zero-shot evaluation across multiple datasets, validate the superior performance and adaptability of FaultGPT in diverse industrial scenarios.
\end{abstract}

\begin{IEEEkeywords}
Industrial Fault Diagnosis, Large Vision Language Models, Large Language Models, Industrial AI, Instruction Tuning, Multimodal Model.
\end{IEEEkeywords}

\section{Introduction}
\label{sec:introduction}
The emergence of large models, including large language models (LLMs) \cite{touvron2023llama,chiang2023vicuna,achiam2023gpt}, vision models \cite{dosovitskiy2020image,radford2021learning}, time-series models \cite{zhou2024one}, and multimodal models based on LLMs \cite{liu2024visual,su2023pandagpt}, has sparked a developmental wave across various academic disciplines. These models, with their billions to trillions of parameters, have been pre-trained on extensive datasets, accumulating vast knowledge and demonstrating unprecedented capabilities in context learning, reasoning, and planning. 
Their application in fields such as industrial robotics \cite{fan2024embodied}, autonomous driving \cite{10529537}, and industrial anomaly detection \cite{gu2024anomalygpt} demonstrate their immense potential in driving innovation and transformation in the industrial sector.

\begin{figure}[t]
    \centering
    \includegraphics[width=0.8\linewidth]{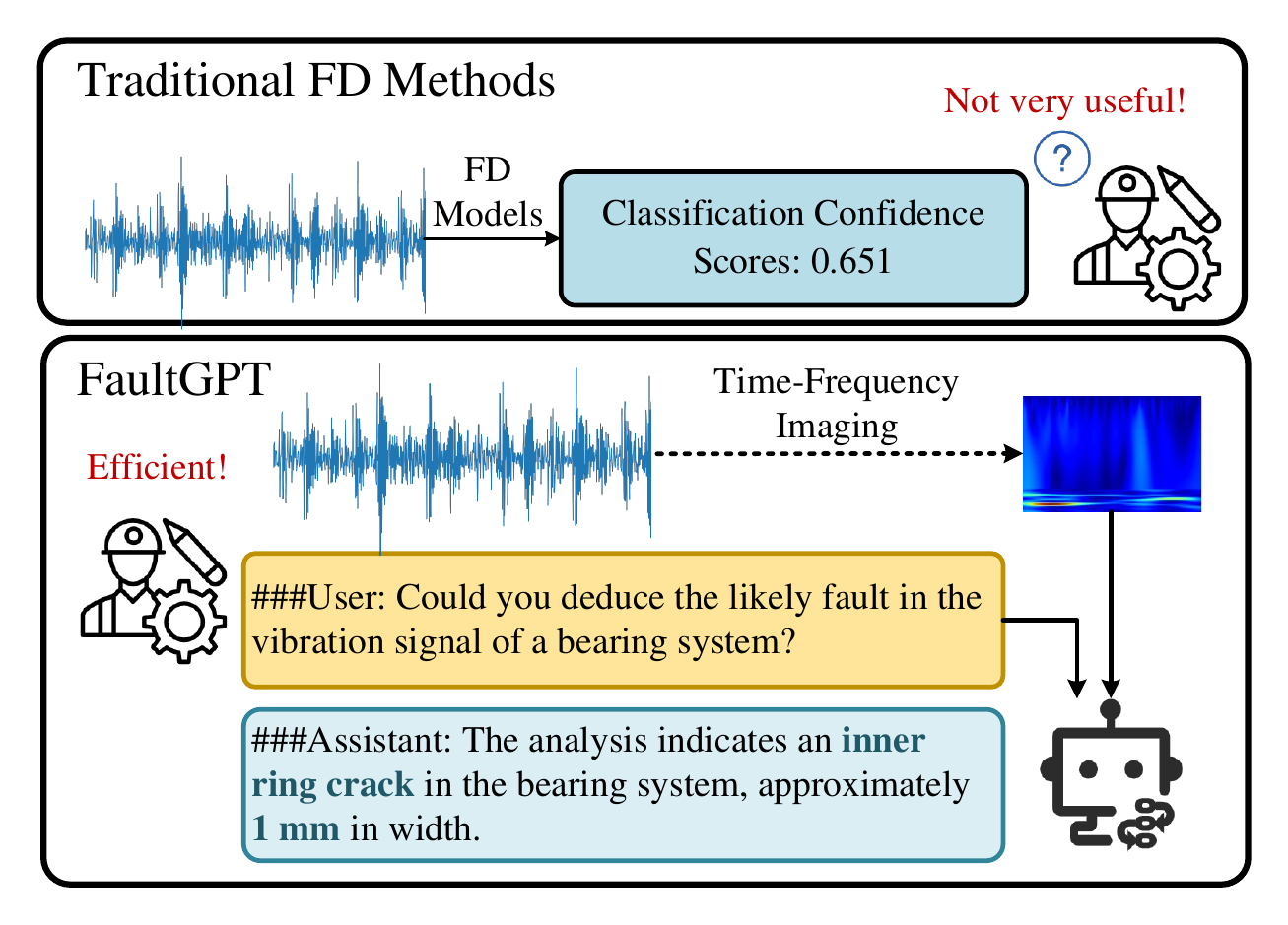}
    \caption{Inference process of FaultGPT compared to traditional fault diagnosis methods.
    }
    \label{fig:fdqa}
    \vspace{-0.15in}
\end{figure}

Fault diagnosis (FD) aims to map the measured signal features to the state of mechanical components \cite{10750065,10609509}. This process involves not only determining the presence of faults but also entails more complex assessments such as fault level, severity, remaining useful life, and the cause of faults.
Traditional deep learning methods, such as those based on CNNs for supervised \cite{9786640,zhang2017new,10102331} or unsupervised learning \cite{10402554}, are typically trained for specific tasks and datasets, limiting their generalizability in dynamic industrial environments. 
In contrast, methods based on pre-trained models \cite{dosovitskiy2020image} can generalize better to specific tasks but require fine-tuning due to the unique nature of mechanical data. 
Recently, time-series models based on LLMs have been introduced as \textit{Tuning Predictors} in the field of mechanical FD, offering a new perspective \cite{zhou2024one}. These models combine pre-trained knowledge with fine-tuning on domain-specific data, demonstrating greater adaptability in dynamic industrial environments.

\begin{figure*}[t]
    \centering
    \includegraphics[width=0.8\linewidth]{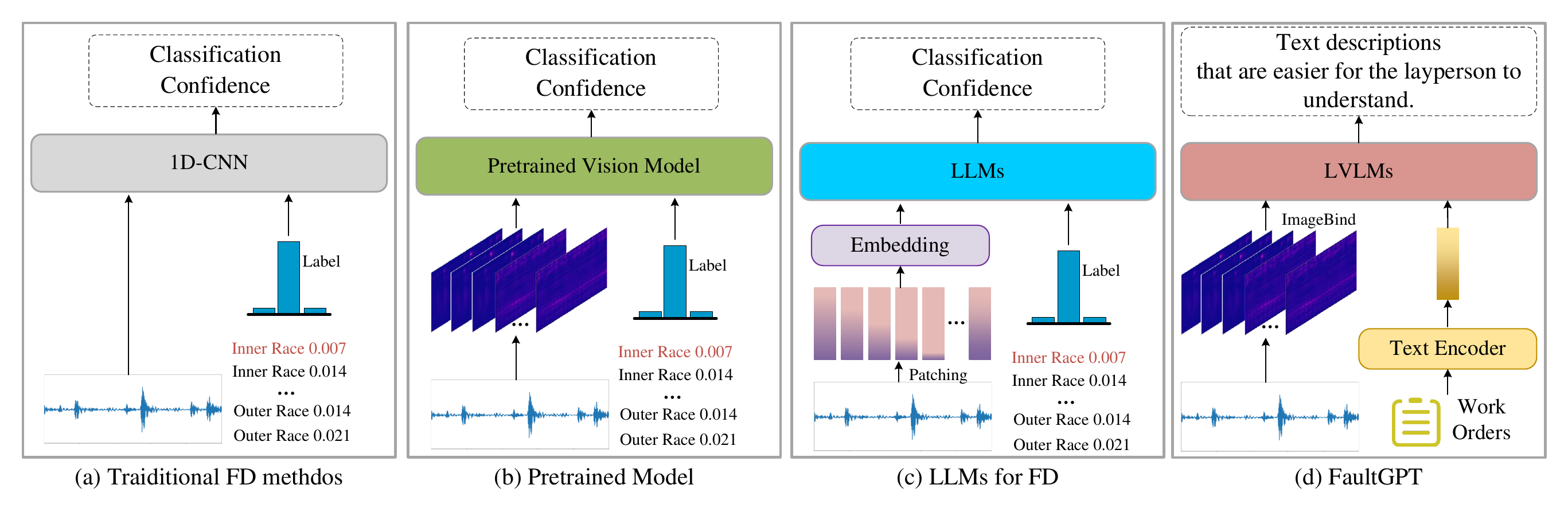}
    \caption{Comparison of previous fault diagnosis methods with our proposed FaultGPT.
    }
    \label{fig:rodemap}
    \vspace{-0.15in}
\end{figure*}

However, the potential of current methods to leverage multimodal data has not been fully exploited \cite{yang2021multiple}. Especially in complex mechanical systems, a single data source often fails to capture fault information comprehensively. Multimodal approaches, integrating different types of data such as acoustic, vibration, and natural language, aim to create a more comprehensive view of mechanical health. For instance, utilizing natural language annotations for weakly supervised optimization of FD models provides new insights for diagnosis systems based on vibration signals \cite{lowenmark2021technical}. \looseness=-1

In this paper, we present FaultGPT, an innovative conversational model that leverages large vision language models (LVLM) to perform end-to-end fault diagnosis question answering (FDQA).
As illustrated in Fig.~\ref{fig:fdqa}, FaultGPT is specifically designed to analyze industrial equipment fault signals by identifying fault types, assessing severity, and locating fault areas with high precision.
Unlike traditional fault diagnosis methods, which often rely on classification confidence scores, FaultGPT integrates data from multiple modalities, including time-frequency images and fault text descriptions, to achieve deep semantic understanding and provide detailed diagnostic reports.
The main novelty and contributions of our work are as follows:

$\bullet$ Our primary contribution is the introduction of FDQA, a novel paradigm for automating the generation of diagnostic reports from vibration signals. This approach goes beyond traditional classification or regression models, which typically offer only limited confidence scores. Instead, FDQA provides an interpretable diagnostic reports, representing a significant advancement in developing sophisticated, user-centric FD systems, enhancing interpretability and supporting more informed decision-making processes.

$\bullet$ Utilizing data collected in the lab and two public datasets, we constructed a comprehensive multimodal instruction dataset that includes vibration time-frequency images, human instructions, and paired diagnostic reports.

$\bullet$ We propose an effective and efficient instruction tuning approach. By leveraging the multi-scale cross-modal image decoder (MCID) to extract fine-grained fault semantics, our method enables LLMs to accurately interpret vibration time-frequency images and generate precise fault diagnosis reports.

$\bullet$ We conducted extensive experiments, including fault diagnosis report generation, few-shot and zero-shot learning across datasets. The results demonstrate the superior performance and transferability of FaultGPT, showcasing its potential for robust fault diagnosis in real-world applications.

\section{Related Works}
\label{sec:related_works}

\subsection{Intelligent Fault Diagnosis}
The development of intelligent FD can be categorized into four distinct research types.
Initial studies predominantly concentrated on acquiring vibration signals from sensors attached to mechanical equipment and devising diagnostic models utilizing one-dimensional 1D-CNNs \cite{zhang2017new,10419797,10571357}, as illustrated in Fig.~\ref{fig:rodemap}(a). 
Despite their effectiveness, these models often suffered from a reliance on limited datasets and the necessity of training from scratch, which poses a considerable challenge in industrial settings where acquiring data is expensive.

As technology advanced, the focus shifted toward leveraging pre-trained models, which reduced dependence on extensive labeled fault data. A notable breakthrough involved transforming vibration signals into time-frequency images \cite{10750065,shao2018highly}, thereby enabling the use of more sophisticated pre-trained models like ResNet \cite{he2016deep} and ViT \cite{dosovitskiy2020image} for mechanical FD, as depicted in Fig.~\ref{fig:rodemap}(b).
Simultaneously, LLMs have demonstrated broad applicability in handling time-series data, particularly excelling in tasks such as classification and prediction \cite{zhou2024one}, offering new avenues for research in this area, as depicted in Fig.~\ref{fig:rodemap}(c).

Despite the progress, generalizing pre-trained models to more complex mechanical systems remains a challenge. Methods relying on single data sources often fail to fully capture the complexity of fault patterns. To address this, multimodal approaches have emerged, integrating different types of data, such as acoustic, vibration, and natural language, to construct a more holistic view of mechanical health monitoring. 
Due to the underutilization of extensive fault and maintenance records in textual form, Natural Language Processing (NLP) techniques are employed to optimize vibration signal-based fault diagnosis models, showcasing the potential of modality fusion in improving diagnostic accuracy \cite{lowenmark2021technical}. In this context, researchers propose the Knowledge Graph-based Question Answering system (KG-MQA) \cite{10367777}, which leverages structured information from the knowledge graph to answer user queries in natural language. Furthermore, researchers construct a large-scale dataset of rotating machinery faults and train the foundational diagnostic model DCNDSC \cite{10520331} on it. This model learns general feature representations and fine-tunes for specific fault diagnosis tasks, thereby improving diagnostic accuracy and robustness.

\begin{figure*}[ht]
  \centering
  \begin{subfigure}[b]{0.20\linewidth}
    \includegraphics[width=\linewidth]{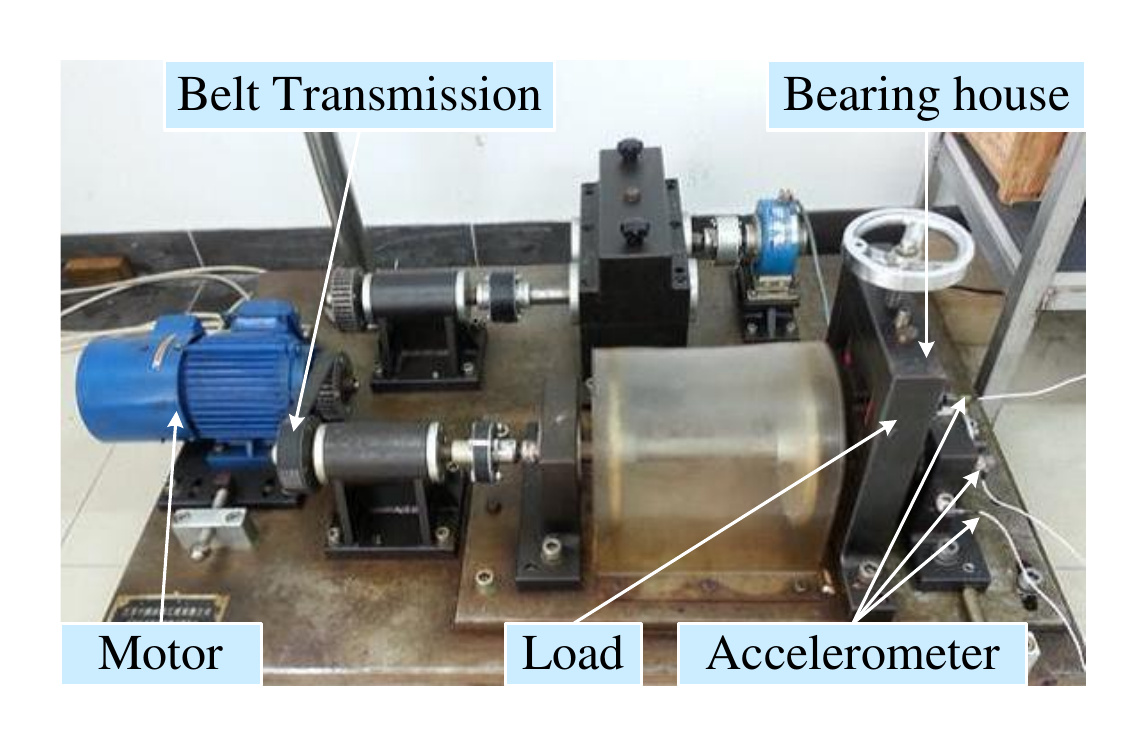}
    \caption{Our vibration signal \\ acquisition equipment.}
    \label{fig:sensors}
  \end{subfigure}
  \begin{subfigure}[b]{0.25\linewidth}
    \includegraphics[width=\linewidth]{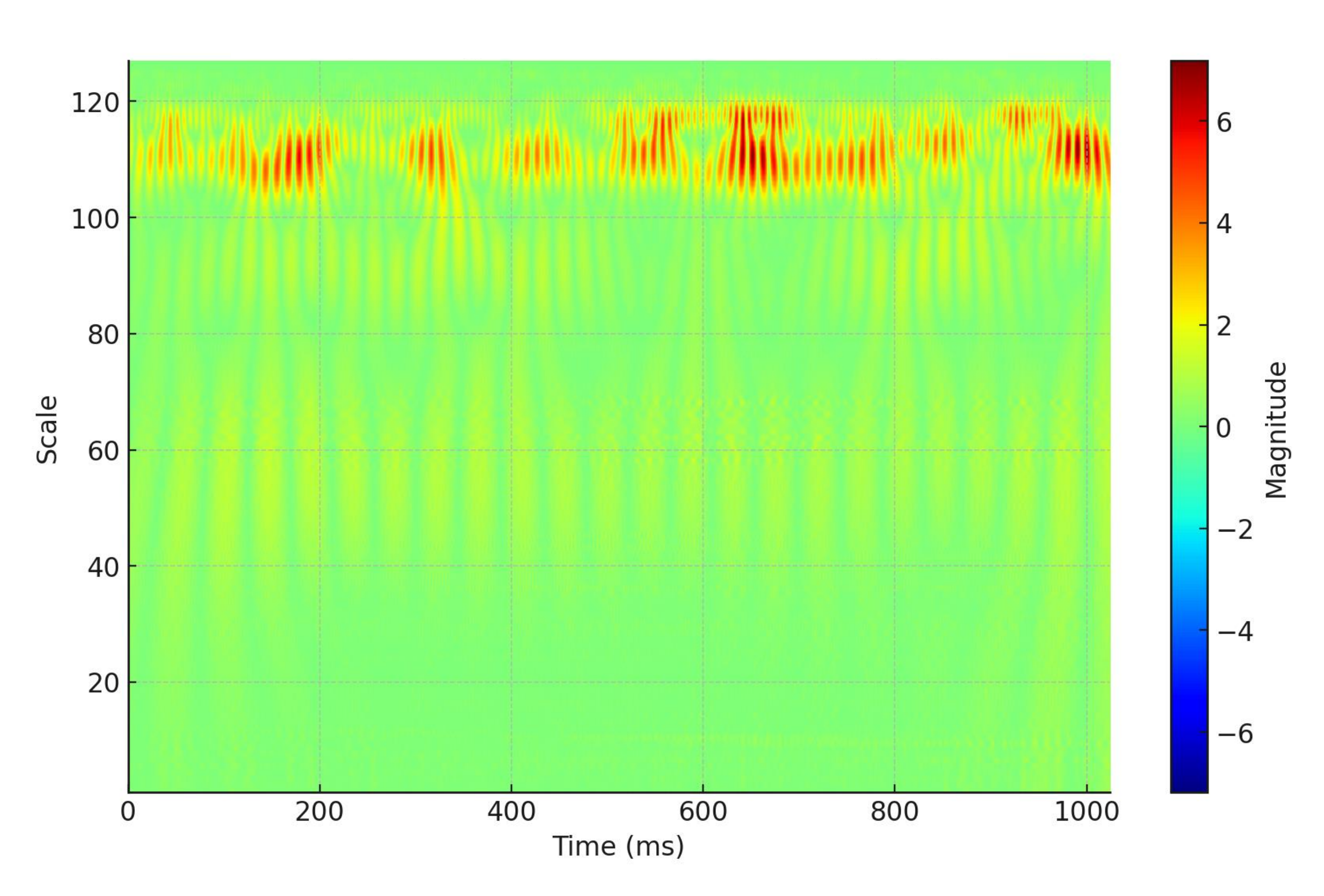}
    \caption{Normal bearing.}
    \label{normal bearing}
  \end{subfigure}
  \begin{subfigure}[b]{0.25\linewidth}
    \includegraphics[width=\linewidth]{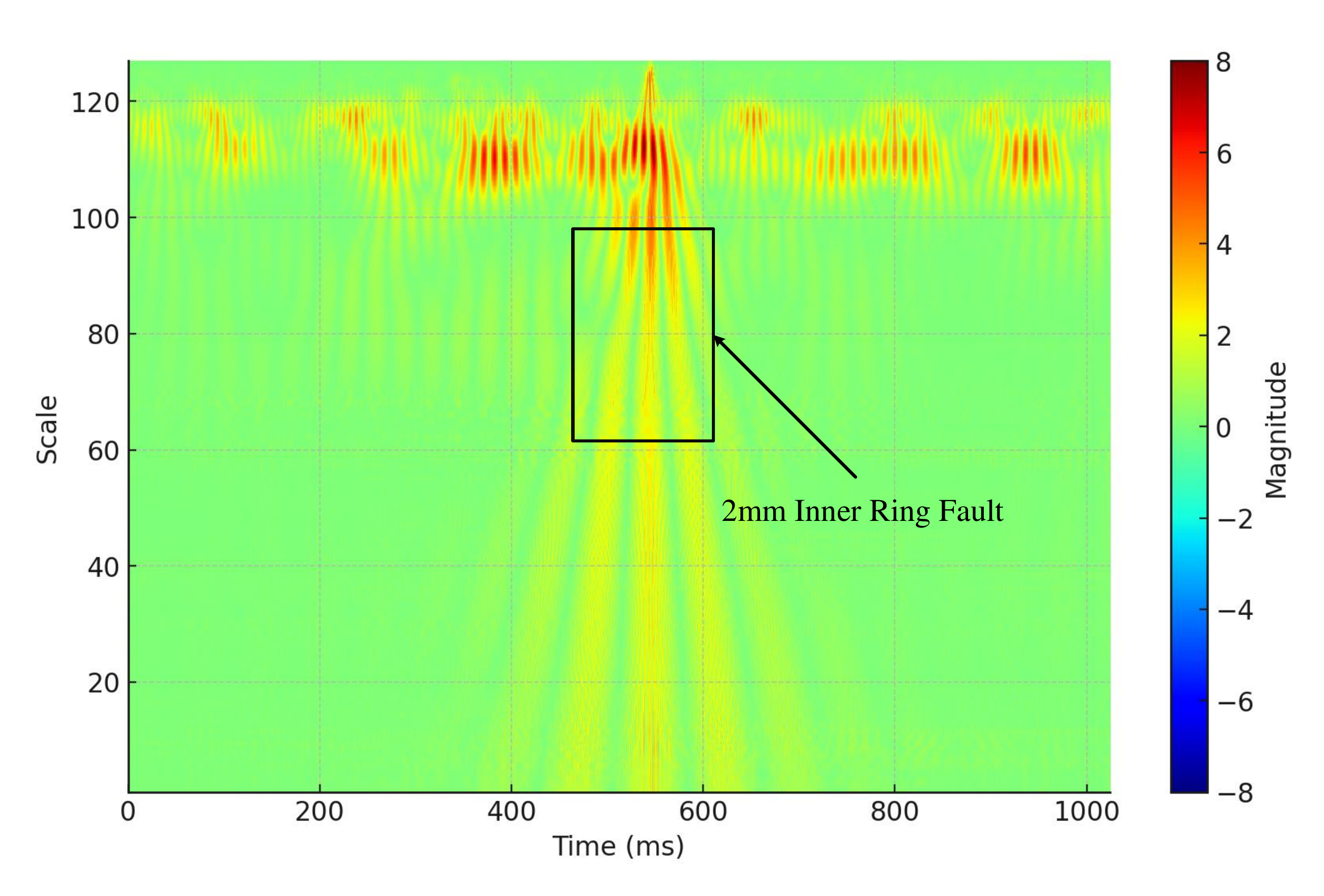}
    \caption{2mm inner ring fault.}
    \label{2mm inner ring fault}
  \end{subfigure}
  \begin{subfigure}[b]{0.22\linewidth}
    \includegraphics[width=\linewidth]{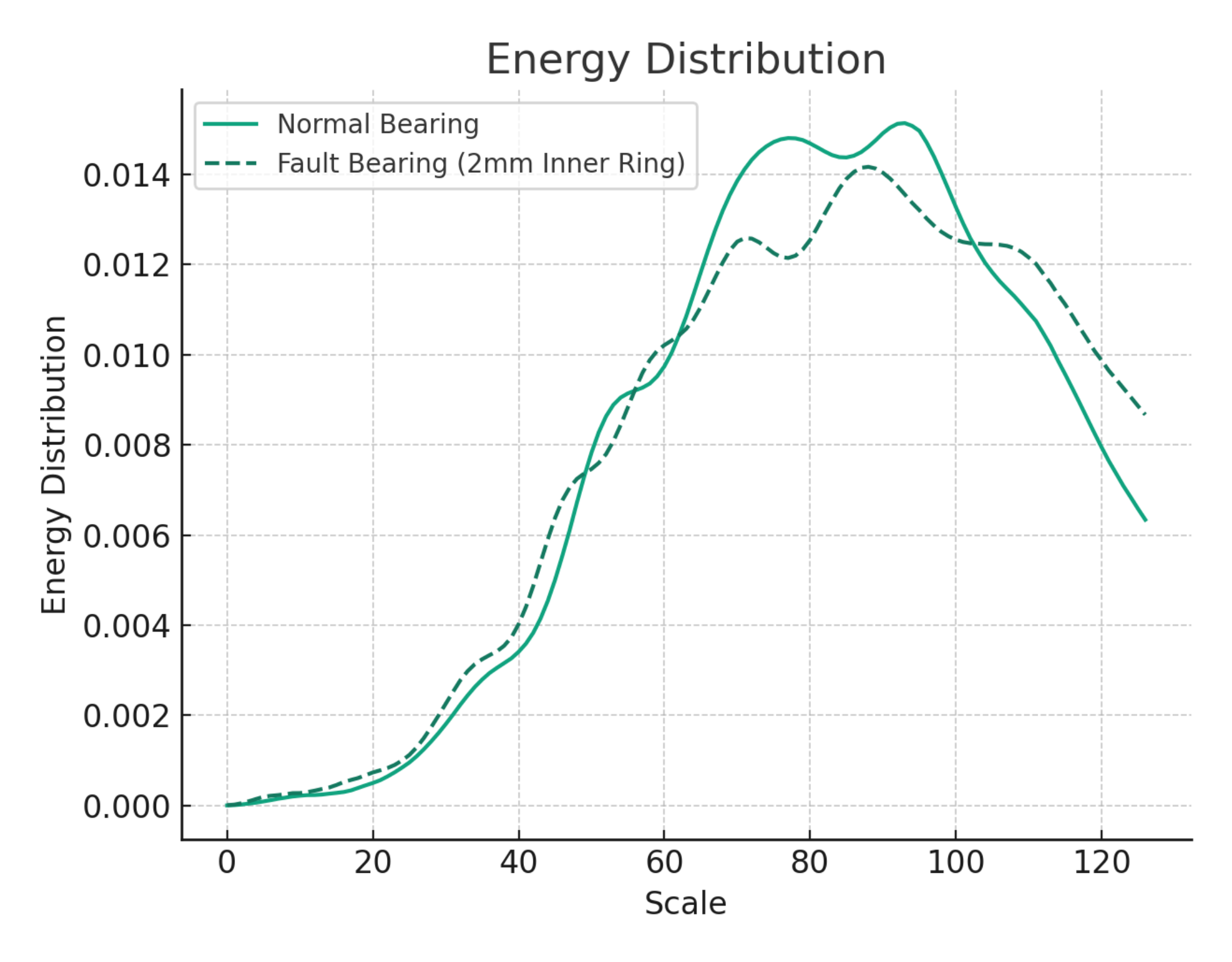}
    \caption{Energy distribution.}
    \label{fig:energy}
  \end{subfigure}
\caption{The time-frequency images and energy characteristics in different bearings.}
\label{fig:observation}
\vspace{-0.15in}
\end{figure*}

\subsection{Multimodal LLMs}
Large models, including LLMs, time-series models, vision models, and large vision language models, represent the forefront of current artificial intelligence research.
LLMs, especially those built on the Transformer architecture, have been pre-trained on extensive text data, accumulating a vast knowledge base and demonstrating exceptional natural language processing capabilities, including text generation, semantic understanding, and sentiment analysis.
These models are categorized into open-source and close-source types. Open-source models like Llama2 \cite{touvron2023llama} and Vicuna\cite{chiang2023vicuna} allow direct access to their architecture and pre-trained weights. In contrast, close-source model such as GPT-4 \cite{achiam2023gpt} are typically accessible via API services, requiring users to craft carefully designed prompts for specific inference tasks.

The multimodal fusion methods can be broadly categorized into two types: (1) \textbf{Standard Cross-attention based Deep Fusion}, which utilizes cross-attention mechanisms within the model to fuse multimodal inputs internally, as in LLAMA-Adapter \cite{gao2023llama}, commonly used in scenarios requiring fine alignment and interaction between modalities. (2) \textbf{Custom Layer based Early Fusion}, where custom modules (such as Q-Former or MLP) directly fuse multimodal data at the input stage. BLIP-2 \cite{li2023blip} performs early fusion using Q-Former, while LLaVA~\cite{liu2024visual} employs an MLP module. Non-tokenized input modalities (such as images or audio) are encoded and then connected to the LLM's text input through these modules, achieving alignment and fusion of multimodal features. In FaultGPT, we adopt the same fusion approach as LLaVA.

Despite these large models achieving remarkable performance in various tasks, significant challenges remain in using vibration signals to generate diagnostic reports: firstly, existing models focus mainly on global feature alignment, struggling to capture local features that represent fault semantics, thus presenting challenges in generating diagnostic reports; secondly, due to the scarcity of fault diagnosis data, there is a lack of relevant benchmarks and datasets.

\section{Methodology}
\label{sec:method}

\subsection{Priliminary}
As shown in Fig.~\ref{fig:observation}(a), accelerometers are typically deployed on machinery to capture vibrational signals during mechanical operations for the purpose of fault diagnosis.
The collected vibrational signals usually constitute a multivariate time series dataset. A vibration signal sample is denoted as \( \mathbf{X}_{\text{vib}} \in \mathbb{R}^{S \times T} \), where \( S \) represents the number of sensors, and \( T \) indicates the length of the signal. 
To facilitate more effective analysis, we transform \( \mathbf{X}_{\text{vib}} \) into its corresponding time-frequency representation, denoted as \( \mathbf{X}_{\text{img}} \), with the detailed transformation process provided in Appendix A.
This transformation enables the mapping of time-series signals to the frequency domain, generating more interpretable image representations. 
Each sample is typically associated with a corresponding maintenance work order \( \mathbf{X}_{\text{text}} \), which is used solely during the training phase,  while the inference phase relies only on the image representation.
The data is collectively represented as a data pair \( \{\mathbf{X}_{\text{img}}, \mathbf{X}_{\text{text}}\} \). 
This formulation is based on the definitions provided in previous research \cite{10750065,lowenmark2021technical}.

\begin{figure*}[t]
    \centering
    \includegraphics[width=0.8\linewidth]{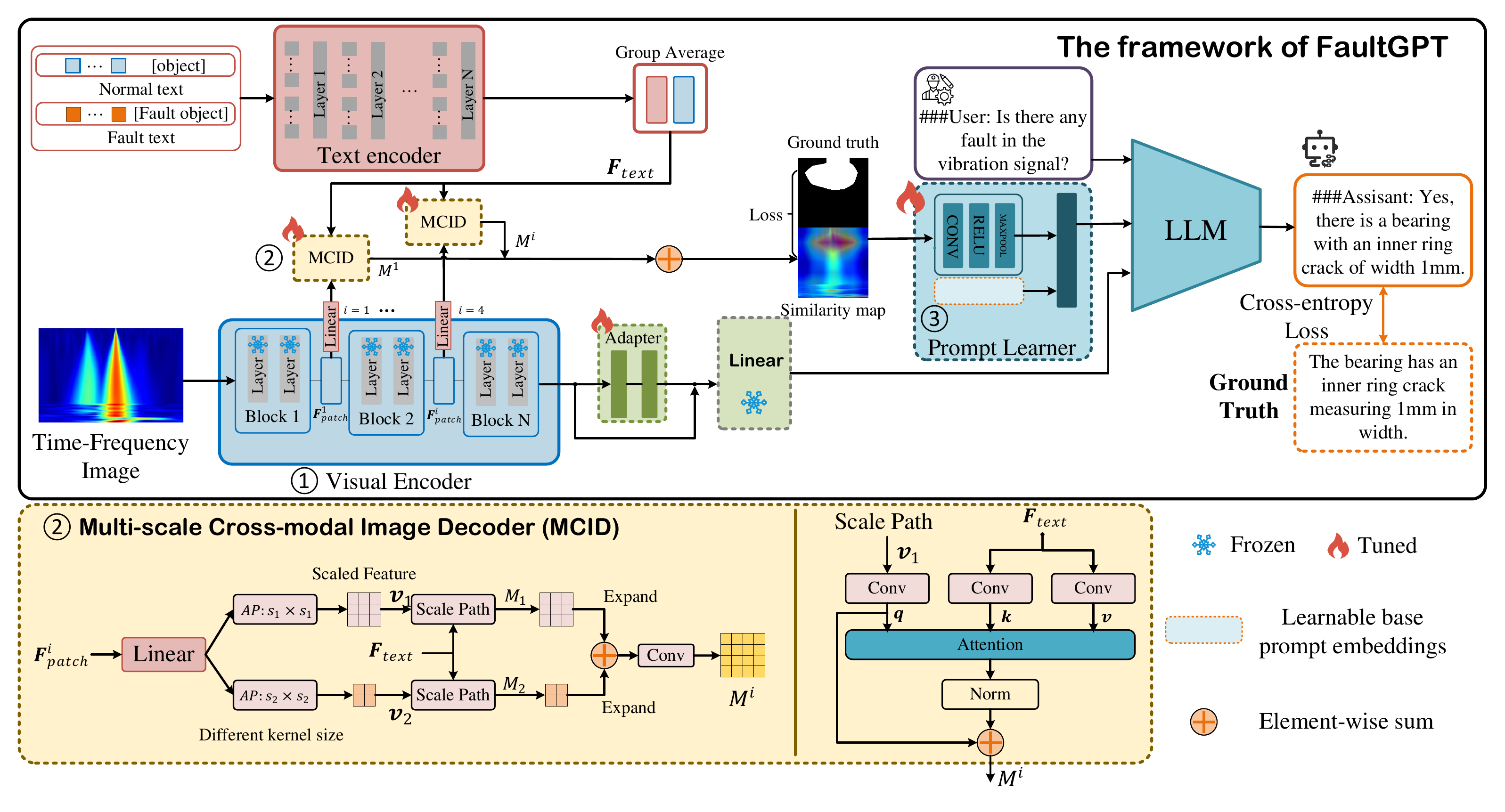}
    \caption{The overall training framework of the proposed \textbf{FaultGPT}. \ding{172} represents the visual encoder, \ding{173} indicates proposed multi-scale cross-modal image decoder, and \ding{174} donates proposed prompt learner.
    }
    \label{fig:framework}
    \vspace{-0.15in}
\end{figure*}

\textbf{Overall Architecture.} As shown in Fig.~\ref{fig:framework}, FaultGPT is a novel conversational FD language-vision model designed to identify fault types, assess fault severity, and accurately locate fault areas, ultimately generating textual descriptions. The architecture consists of three key components:
\begin{itemize}
    \item \textbf{\ding{172} Visual Encoder:} This component encodes the input time-frequency images and projects the visual features into an embedding space compatible with the language model, ensuring effective alignment between visual and linguistic features.
    
    \item \textbf{\ding{173} Multi-Scale Cross-Modal Image Decoder (MCID):} The MCID is responsible for extracting more detailed fault features, especially focusing on local fault information. This enables the model to capture finer details of the vibration signals, improving fault diagnosis accuracy.
    
    \item \textbf{\ding{174} Prompt Learner:} The Prompt Learner integrates the fine-grained fault semantics extracted by the MCID with the learnable base prompt embeddings. This fusion enables the model to generate more accurate and context-aware fault descriptions by combining the localized fault information with additional semantic details. 
\end{itemize}

This architecture allows for seamless interaction between visual and linguistic features within a unified space, enhancing the model's performance in generating relevant and precise textual outputs for fault diagnosis tasks.

\subsection{Components of FaultGPT }
\subsubsection{Visual Encoder}
In the visual encoder of FaultGPT, the effective extraction of time-frequency image features is a crucial step to ensure efficient interaction with semantic embeddings in the LLM backbone. We utilize the pre-trained CLIP \cite{radford2021learning} model (ViT-L/14) as the foundation for the visual encoder, which has demonstrated exceptional performance in multimodal tasks, such as image and text alignment.

To further enhance the model's adaptability for fault diagnosis tasks, we introduce an Adapter \cite{houlsby2019parameter} module within the visual encoder. The purpose of incorporating the Adapter is to enable fine-tuning with a small number of parameters while keeping the weights of the CLIP backbone frozen. This approach allows us to leverage the generalization capabilities of the pre-trained model while simultaneously adapting to the specific needs of the task, all while reducing computational overhead and model complexity.

The visual encoder uses a linear projection layer to map high-dimensional image features into a lower-dimensional representation space that aligns with the semantic embeddings in the LLM. This projection process can be expressed as:
\begin{equation}
    \mathbf{H}_{\text{img}} = \mathcal{P}(\mathcal{F}(\mathbf{X}_{\text{img}})),
\end{equation}
where \(\mathbf{H}_{\text{img}}\) represents the embedding vector extracted from the image, \(\mathcal{P}\) is the linear projection function, \(\mathcal{F}\) is the feature extraction function from the time-frequency image, and \(\mathbf{X}_{\text{img}}\) is the input time-frequency image. 

\textbf{Remark}: To enable efficient multimodal fusion, we adopt the ImageBind \cite{girdhar2023imagebind}, which projects both visual and textual features into a shared representation space. This alignment ensures effective interaction between the time-frequency image features and the semantic embeddings in the LLM, facilitating seamless cross-modal comparisons.

\subsubsection{Multi-scale Cross-modal Image Decoder}
Observation of the time-frequency image of bearings from Fig.~\ref{fig:observation}(b)-\ref{fig:observation}(c) indicates that the vibration signals of a normal bearing exhibit a relatively random distribution, while those of a faulty bearing show distinct impulsive characteristics at specific time points. In addition, the energy distribution, as shown in Fig.~\ref{fig:observation}(d), further reveals the differences in energy distribution between normal and faulty bearings across various scales, particularly highlighting the substantially higher energy of faulty bearings at certain scales.

Although we utilize the CLIP model's visual encoder for extracting the global visual features, especially object semantics, it is insufficient to fully capture the fine-grained fault-related semantics within time-frequency images \cite{zhou2023anomalyclip}. Thus, we introduce the MCID, which is designed to handle cross-modal data and extract detailed local fault semantics. 
MCID effectively captures the detailed correspondence between fault textual descriptions and the fine-grained, local fault regions of time-frequency images, facilitating the comprehensive understanding of fault information by grasping image features across multiple scales.

Inspired by ClipSAM \cite{li2024clipsam}, we adopt the \textit{Scale Path} to capture global features of images across different scales for a thorough understanding of faults. Let \( \mathbf{F}_{\text{text}} \in \mathbb{R}^{C_{\text{text}} \times 2} \) denote the text feature vector, which is obtained on the same lines as winCLIP \cite{jeong2023winclip}. The image features are extracted from different stages of the image encoder, denoted by \( \mathbf{F}_{\text{patch}}^i \), where \( i \) indicates the stage number.

The image features are initially projected onto \( \hat{\mathbf{F}}_{\text{patch}} \in \mathbb{R}^{H \times W \times C_{\text{text}}} \) via a linear layer. Subsequently, two average pooling layers of different sizes (kernel sizes \( s_1 \) and \( s_2 \)) are applied to capture fault features at multiple scales, represented as: \looseness=-1
\begin{align}
\mathbf{v}_{1} &= \text{conv}_{3 \times 3} (\text{Avg\_Pool}_{s_1 \times s_1} (\hat{\mathbf{F}}_{\text{patch}})), \\
\mathbf{v}_{2} &= \text{conv}_{3 \times 3} (\text{Avg\_Pool}_{s_2 \times s_2} (\hat{\mathbf{F}}_{\text{patch}})),
\end{align}
where \( \mathbf{v}_{1} \) and \( \mathbf{v}_{2} \) denote fault features at two distinct scales. Subsequently, text features undergo convolutional processing to yield \( \mathbf{F}_{\text{text},1}^k, \mathbf{F}_{\text{text},1}^v \in \mathbb{R}^{C_{\text{text}} \times 2} \), which interact with \( \mathbf{v}_{1} \) and \( \mathbf{v}_{2} \) to produce pixel-representing features \( \mathbf{M}_{1} \) and \( \mathbf{M}_{2} \).
In what follows, we use \( \mathbf{v}_{1} \) as an example, and \( \mathbf{v}_{2} \) is similar.
An attention mechanism is employed, designating \( \mathbf{v}_{1} \) as the query, \( \mathbf{F}_{\text{text},1}^k \) as the key, and \( \mathbf{F}_{\text{text},1}^v \) as the value, to amplify the correlation between textual and time-frequency image. The features \( \mathbf{M}_{1} \) and \( \mathbf{M}_{2} \) are then resized to their original dimensions and merged via linear interpolation, described by the equation:
\begin{equation}
\mathbf{M}^{i} = \text{conv}_{3 \times 3} (\mathcal{B}(\mathbf{M}_{1}) + \mathcal{B}(\mathbf{M}_{2})),
\end{equation}
where \( \mathcal{B} \) represents the bilinear interpolation layer and \(\mathbf{M}^{i}\) indicates the merged feature of intermediate stage $i$ derived from multiple scales.

To learn both global and granular fault semantics, we integrate all merged features \(\mathbf{M}^{i}\) to provide more detailed local fault feature insights.
Formally, let $\mathcal{I}$ be the set of intermediate layers used.
The final local fault features \( \mathbf{M} \) is obtained by the following equation,
\begin{equation} \label{mlp}
    \mathbf{M} = \frac{1}{|\mathcal{I}|} \sum_{i\in\mathcal{I}} \text{MLP}(\text{ReLU}(\mathbf{M}^{i})),
\end{equation}
where we employ a Multi-Layer Perceptron for rough localization of the fault information. In practice, we divide the image encoder into four stages to extract intermediate patch-level features at each stage, as shown in Fig.~\ref{fig:framework}. Equation \eqref{mlp} enables the accurate extraction of faults features in time-frequency images, which is significant for achieving industrial-level accurate fault fault.

Since MCID contains learnable parameters, we leverage Grad-CAM to construct supervised learning objectives. The specific construction process is detailed in Appendix B.

\subsubsection{Prompt Learner}
In FaultGPT, the role of the prompt learner is to align the fault features extracted from time-frequency images with corresponding textual outputs, ensuring that the model generates accurate fault diagnosis descriptions. The fault features are extracted by the MCID and represented as \( \mathbf{M} \in \mathbb{R}^{H \times W} \), where \( H \) and \( W \) denote the height and width of the feature map, respectively.

The prompt learner transforms these fault features \( \mathbf{M} \) into prompt embeddings interpretable by LLMs, thereby providing additional semantic information to ensure precise alignment between visual features and language generation. The prompt learner comprises two main components: learnable base prompt embeddings \( E_{\text{base}} \) and decoded prompt embeddings \( E_{\text{dec}} \).

First, the prompt learner contains a set of learnable base prompt embeddings \( E_{\text{base}} \in \mathbb{R}^{n_1 \times C_{\text{emb}}} \), where \( n_1 \) represents the number of base embeddings and \( C_{\text{emb}} \) denotes the embedding dimension (experimentally set to 4096). These base prompt embeddings are independent of the fault features and provide additional context information, enhancing the diagnostic capabilities of the model.

Second, the fault features \( \mathbf{M} \) extracted by the MCID are transformed through a convolutional neural network into \( n_2 \) decoded prompt embeddings \( E_{\text{dec}} \in \mathbb{R}^{n_2 \times C_{\text{emb}}} \), where \( n_2 \) denotes the number of new embeddings generated (with \( n_1 = n_2 = 9 \) in our experiments). These decoded prompt embeddings, along with the base prompt embeddings, form a complete set of prompt embeddings \( E_{\text{prompt}} \in \mathbb{R}^{(n_1+n_2) \times C_{\text{emb}}} \).

As illustrated in Fig.~\ref{fig:framework}, \( E_{\text{prompt}} \) is combined with the image embeddings before being input into the LLM, which enhances the alignment accuracy between fault features and language generation. The design of the prompt learner allows for the effective fusion of visual and language features in a shared space, thereby improving the precision of fault diagnosis reports generated by the model. 
The specific implementation details of the prompt learner can be found in Appendix D.

\subsection{Instruction Tuning}
\subsubsection{Multimodal Feature Alignment}
\label{llms}
We combine the time-frequency image features \( \mathbf{H}_{\text{img}} \) with the prompt embeddings \( \mathbf{E}_{\text{prompt}} \) processed by the prompt learner, which are then fed into the LLM to guide the next-token prediction task. This method aligns both visual and textual features, enhancing multimodal interaction. The corresponding equation is similar to the previous method but with modified notation:
\begin{equation}
    p(\mathbf{X}_t | \mathbf{H}_{\text{img}}, \mathbf{E}_{\text{prompt}}) = \prod_{i=j}^{L} p_{\theta} (\mathbf{x}_{t,i} | \mathbf{H}_{\text{img}}, \mathbf{E}_{\text{prompt}}, \mathbf{X}_{t, <i}),
\end{equation}
where \( j \) is the starting index after \texttt{<assistant>}, \( \theta \) represents the trainable parameters of the prompt learner and the LLM within FaultGPT, and \( \mathbf{X}_{t, <i} \) denotes all the response tokens generated prior to the current token \( \mathbf{x}_{t,i} \). This approach ensures effective fusion of visual and linguistic features, allowing the model to capture fine-grained fault information during text generation.

\subsubsection{Loss Function}
Our model's architecture, incorporating both a MCID and a prompt learner, requires multiple loss functions to tackle distinct challenges during training, thereby enhancing precision and adaptability.

\textbf{Cross-entropy Loss.}
The cross-entropy loss is crucial for models generating probabilistic predictions, aiming to align the predicted sequence of tokens with the target sequence, thus fine-tuning the model's predictive accuracy. The cross-entropy loss is defined as:
\begin{equation}
    L_{\text{ce}} = -\sum_{c=1}^{N} \textbf{y}_c \log(\hat{\textbf{y}}_c),
\end{equation}
where \(N\) represents the number of tokens, \(\textbf{y}_c\) is the true class label of the \(c\)-th token, and \(\hat{\textbf{y}}_c\) is the predicted probability for that token.

\textbf{Focal Loss.}
Tailored to address the challenge of class imbalance in image data, focal loss modifies the standard cross-entropy loss to enhance the model’s focus on hard-to-classify pixels, improving sensitivity to less represented classes. The focal loss is defined as:
\begin{equation}
    L_{\text{focal}} = -\frac{1}{M} \sum_{c=1}^{M} (1 - \hat{\textbf{y}}_c)^\gamma \log(\hat{\textbf{y}}_c),
\end{equation}
where \( M \) is the total number of pixels, \( \hat{\textbf{y}}_c \) represents the predicted probability that the \(c\)-th pixel belongs to the target class, and \( \gamma \) is a tunable parameter that scales the contribution of each pixel based on its classification difficulty, typically set to 2. \looseness=-1

\textbf{Dice Loss.}
Primarily utilized in semantic segmentation, dice loss evaluates the similarity between the predicted and actual segmentation masks, particularly beneficial for handling class imbalances within images:
\begin{equation}
    L_{\text{dice}} = 1 - \frac{2 \sum_{c=1}^{M} \hat{\textbf{y}}_c \textbf{y}_c}{\sum_{c=1}^{M} \hat{\textbf{y}}_c^2 + \sum_{c=1}^{M} \textbf{y}_c^2},
\end{equation}
where \( \textbf{y}_c \) is the actual label of the \(c\)-th pixel, and \( \hat{\textbf{y}}_c \) is the predicted probability of the \(c\)-th pixel being part of the target class.
To foster a balanced training approach, the overall loss combines these individual losses, each modulated by a specific coefficient:
\begin{equation}
    L = \alpha L_{\text{ce}} + \beta L_{\text{focal}} + \delta L_{\text{dice}},
\end{equation}
where \(\alpha\), \(\beta\), and \(\delta\) act as balancing coefficients for the composite loss function, typically initialized to unity for equal weighting initially.

\section{FDQA Instruction-following Data}
\label{sec:it_data}
In this section, we first introduce the datasets used in this study, followed by the two categories of textual input formats: the first describes the input time-frequency images, while the second presents the instruction-tuning data format. Lastly, we detail the evaluation metrics used to assess model performance.

\label{Normal and Fault Texts}
\begin{table}[!t]
    \centering
    \caption{Detailed text label for every category in the SCUT-FD dataset.}
    \label{tab:fault_text}
    \begin{tabular}{@{}cp{6cm}@{}}
      \toprule
      \textbf{Class} & \textbf{Time-Frequency Image Description (label)} \\
      \midrule
      Normal bearing & This is a time-frequency image of a bearing for fault diagnosis, which is normal and fault-free. \\
      \midrule
      Faulty bearing & This is a time-frequency image of a bearing with an \{fault\_type\} fault of width \{fault\_width\}, depth \{fault\_depth\}. \newline 
      e.g., an inner ring fault with width 1mm, depth 0.5mm. \\
      \bottomrule
    \end{tabular}
    \vspace{-0.15in}
\end{table}

\subsection{FDQA Instruction Data Curation}
\label{sec:datasets}
\textbf{Datasets.} This research utilizes three large-scale bearing fault datasets: the CWRU \cite{smith2015rolling}, the SCUT-FD and Ottawa \cite{huang2018bearing} bearing dataset. 
The CWRU is extensively used in the study of FD. It includes vibration signals of bearings under various operating conditions with fault types encompassing inner ring, outer ring, and rolling elements, with fault sizes ranging from 0.007 inches to 0.021 inches.
SCUT-FD \cite{li2020novel} is a dataset of bearing FD collected in the laboratory, classified by fault location and size. The collection equipment is shown in Fig.~\ref{fig:observation}(a). We use a total of 9 sensors to collect mechanical vibration signals.
In the Ottawa dataset \cite{huang2018bearing}, the health condition of bearings is categorized into three types: normal, inner ring fault, and outer ring fault. The operating speed conditions include four scenarios, such as gradually increasing speed. This dataset offers high-quality foundational data for studying bearing fault diagnosis and performance under varying speed conditions, supporting effective fault detection across different scenarios.

In this section, we introduce the two types of textual input formats used in this paper, which consist of two categories: the first type is used to describe the input time-frequency image, and the second type is used to query specific fault-related issues.

\subsubsection{Normal and Fault Texts}
The Normal and Fault Texts refer to the textual labels used to describe the time-frequency images. These descriptions provide crucial information about the condition of the bearings, whether they are normal or faulty, and include specific attributes such as fault type, width, and depth.
To generate textual descriptions for both normal and faulty conditions, we employ a synthetic prompting method aligned with CLIP’s text input templates, facilitating easier multimodal alignment. The template used is: “This is a time-frequency image of a bearing with a \{fault\_type\} fault of width \{fault\_width\}, depth \{fault\_depth\}.” Here, \{fault\_type\} can represent an inner or outer ring fault, \{fault\_width\} can be 1mm or 2mm, and \{fault\_depth\} can be 0.5mm or 1mm, among others. By substituting these placeholders, we generate detailed descriptions. Table~\ref{tab:fault_text} provides examples for both normal and faulty samples. This approach improves clarity and aligns with CLIP \cite{radford2021learning}, enhancing multimodal learning and fault recognition accuracy.

\label{instruction-following data}
\begin{figure}[t]
    \centering
    \includegraphics[width=0.9\linewidth]{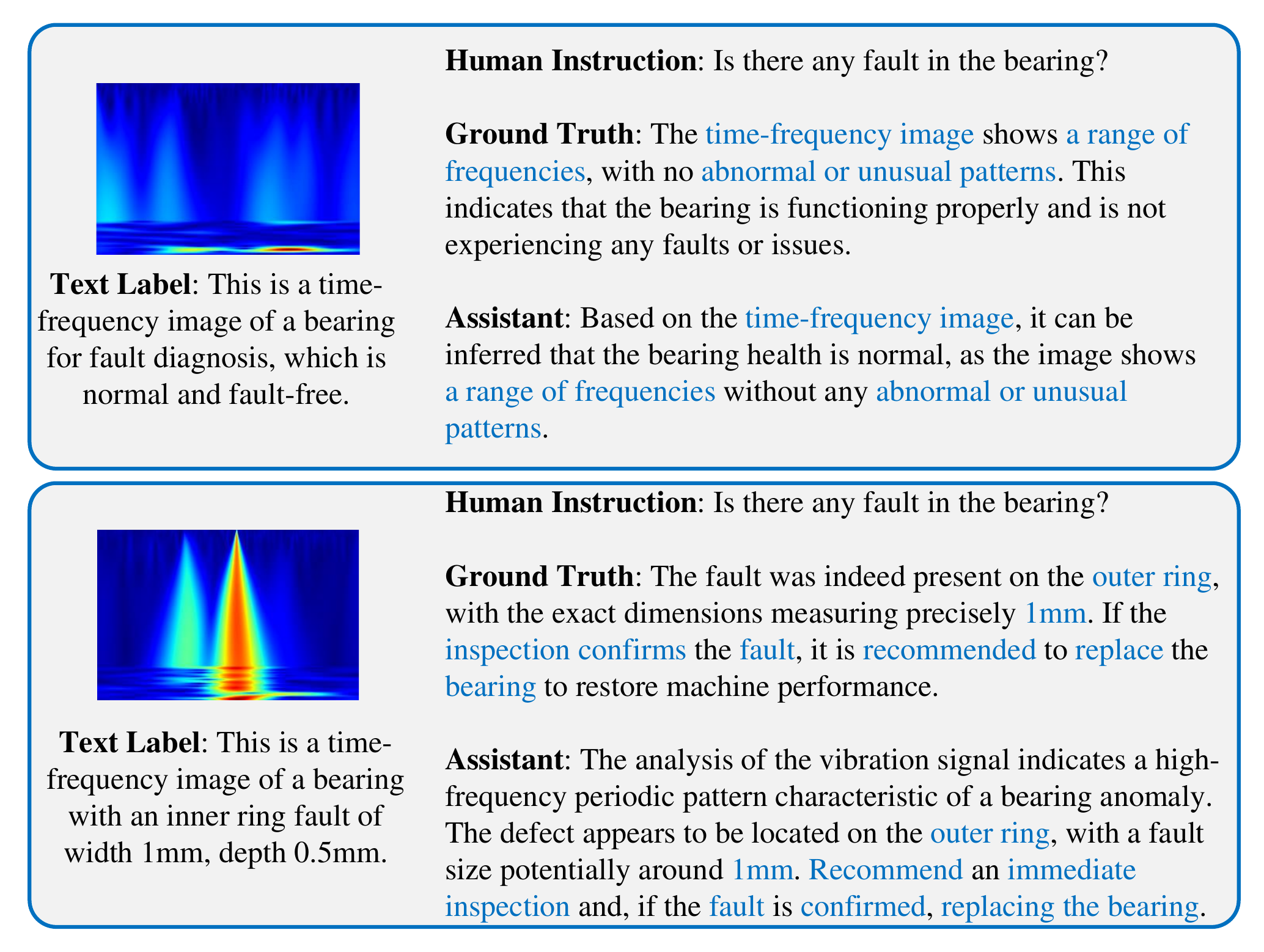}
    \caption{Example of fault diagnosis instruction data.}
    \label{fig:examples}
    \vspace{-0.15in}
\end{figure}

\subsubsection{Instruction Data Format}
The second type of text queries whether the time-frequency image corresponds to a fault, i.e., ``Is there a fault in the bearing shown in this time-frequency image?” The LVLM first responds to whether faults are present. If a fault is detected, the model continues by specifying the location and severity of the fault.

In FDQA, the objective of instruction-tuned data is to equip the language model with the ability to process multimodal inputs, such as image features and human instructions, in order to generate the expected fault diagnosis responses.
In our dataset, we simulate an interactive fault diagnosis task between a human and a machine. The input includes time-frequency images and user instructions. The format is as follows:

\lstset{
  basicstyle=\ttfamily\footnotesize,
  breaklines=true,         
  frame=none                          
}
\begin{lstlisting}
### Human: <Img> {Image_Features} </Img> {Prompt Text}
### Assistant: {Response Text}
\end{lstlisting}

The terms are defined as follows: \textbf{\{Image\_Features\}}: Represents the embedded features of the image, extracted by a deep neural network from the time-frequency image. \textbf{\{Prompt Text\}}: Represents the user's command or question (e.g., ``Please diagnose the fault type in this image."). \textbf{\{Response Text\}}: Represents the response generated by the model (e.g., ``The bearing inner ring has a fault with a width of 1mm and a depth of 0.5mm.").

Fig.~\ref{fig:examples} shows demonstrations of the FDQA instruction-following data. We highlight the consistent information between the generated answers and the ground truths with blue color. \looseness=-1

\subsection{Evaluation Metrics}
We evaluate the model's performance using eight metrics: Accuracy, BLEU 1-4, ROUGE-L, CIDEr-D, and Match. Accuracy measures the proportion of predictions that exactly match the ground truth, reflecting the model's overall performance in classification tasks. BLEU evaluates the quality of machine-generated translations, focusing on both accuracy and fluency. ROUGE-L assesses the fluency and structural coherence of generated sentences. CIDEr-D evaluates the relevance and uniqueness of the generated FD reports against a set of reference candidates. Match, on the other hand, compares the generated results with the ground truth, ensuring that key details such as fault type and fault size are correctly aligned between the generated report and the actual condition.

\section{Experiments and Evaluation}
In this section, we evaluate four open-source LVLM using the FDQA instruction-following datasets, which were constructed as detailed in Section~\ref{sec:it_data}. Additionally, we provide a comprehensive evaluation of scalability and instruction tuning, alongside case studies that demonstrate the generation of fault diagnosis reports.

\subsection{Experimental Setting}
To validate our proposed FaultGPT from various perspectives and levels, we have selecte the following models as baselines for comparative research:

\subsubsection{Models}
Unlike existing fault diagnosis methods, FaultGPT focuses on fault Q\&A, generating diagnostic reports directly from raw data through an interactive design. Therefore, we primarily choose Decoder-only Transformers as comparison methods, as these models are commonly used for generation, inference, and decision-making tasks.
We use 3 LLMs based on the \texttt{peft3} library to construct the multimodal FDQA model described in Section~\ref{llms}. These models include GPT-Neo, Mistral \cite{jiang2023mistral} and LLaMA-2 \cite{touvron2023llama} along with a small pre-trained language model (GPT2-Large \cite{radford2019language}) as founamental baselines.

Detailed experimental settings and configurations can be found in Appendix C.

\subsection{Quality Evaluation}
\label{Quality Evaluation}

\begin{table*}[]
\centering
\caption{FaultGPT performance metrics on various datasets.}
\label{Quality_Evaluation}
\begin{tabular}{l|c|c|cccccc|c}
\toprule
\multirow{2}{*}{Datasets} & \multirow{2}{*}{LLMs} & \multicolumn{1}{l|}{\multirow{2}{*}{Accuracy $\uparrow$}} & \multicolumn{6}{c|}{Language $\uparrow$}                                                                                                                                                                           & \multicolumn{1}{l}{\multirow{2}{*}{Match $\uparrow$}} \\
                          &                         & \multicolumn{1}{l|}{}                                     & Bleu\_1                              & Bleu\_2                              & Bleu\_3                              & Bleu\_4                              & ROUGE\_L                             & CIDEr           & \multicolumn{1}{l}{}                                  \\ \midrule 
\multirow{4}{*}{CWRU}     & GPT2-Large              & 0.5490                                                    & \multicolumn{1}{c|}{0.5761}          & \multicolumn{1}{c|}{0.5271}          & \multicolumn{1}{c|}{0.4562}          & \multicolumn{1}{c|}{0.4215}          & \multicolumn{1}{c|}{0.5223}          & 3.7021          & 38.2512                                               \\
                          & GPT-Neo                 & 0.5886                                                    & \multicolumn{1}{c|}{0.6312}          & \multicolumn{1}{c|}{0.5793}          & \multicolumn{1}{c|}{0.5344}          & \multicolumn{1}{c|}{0.4895}          & \multicolumn{1}{c|}{0.6896}          & 4.8114          & 40.7342                                               \\
                          & Mistral                 & 0.7105                                                    & \multicolumn{1}{c|}{0.8867}          & \multicolumn{1}{c|}{0.8640}          & \multicolumn{1}{c|}{0.8422}          & \multicolumn{1}{c|}{0.8202}          & \multicolumn{1}{c|}{0.8875}          & 7.3282          & 86.3968                                               \\
                          & LLaMA-2                 & \textbf{0.7423}                                           & \multicolumn{1}{c|}{\textbf{0.9011}} & \multicolumn{1}{c|}{\textbf{0.8812}} & \multicolumn{1}{c|}{\textbf{0.8607}} & \multicolumn{1}{c|}{\textbf{0.8397}} & \multicolumn{1}{c|}{\textbf{0.8976}} & \textbf{7.8805} & \textbf{89.7032}                                      \\ \midrule \midrule
\multirow{4}{*}{Ottawa}     & GPT2-Large              & 0.4332                                                    & \multicolumn{1}{c|}{0.3292}          & \multicolumn{1}{c|}{0.2788}          & \multicolumn{1}{c|}{0.2543}          & \multicolumn{1}{c|}{0.2327}          & \multicolumn{1}{c|}{0.3914}          & 2.5216          & 35.8701                                               \\
                          & GPT-Neo                 & 0.5275                                                    & \multicolumn{1}{c|}{0.4748}          & \multicolumn{1}{c|}{0.4491}          & \multicolumn{1}{c|}{0.3987}          & \multicolumn{1}{c|}{0.3732}          & \multicolumn{1}{c|}{0.4863}          & 3.7025          & 39.9880                                               \\
                          & Mistral                 & \textbf{0.7518}                                           & \multicolumn{1}{c|}{0.7787}          & \multicolumn{1}{c|}{\textbf{0.7317}} & \multicolumn{1}{c|}{\textbf{0.6741}} & \multicolumn{1}{c|}{0.6059}          & \multicolumn{1}{c|}{0.7249}          & 4.2033          & 67.6428                                               \\
                          & LLaMA-2                 & 0.7339                                                    & \multicolumn{1}{c|}{\textbf{0.7940}} & \multicolumn{1}{c|}{0.7176}          & \multicolumn{1}{c|}{0.6608}          & \multicolumn{1}{c|}{\textbf{0.6185}} & \multicolumn{1}{c|}{\textbf{0.7463}} & \textbf{4.2746} & \textbf{67.9204}                                      \\ \midrule \midrule
\multirow{4}{*}{SCUT-FD}  & GPT2-Large              & 0.5616                                                    & \multicolumn{1}{c|}{0.4378}          & \multicolumn{1}{c|}{0.3951}          & \multicolumn{1}{c|}{0.3553}          & \multicolumn{1}{c|}{0.3206}          & \multicolumn{1}{c|}{0.4815}          & 3.2514          & 0.5275                                                \\
                          & GPT-Neo                 & 0.6132                                                    & \multicolumn{1}{c|}{0.4855}          & \multicolumn{1}{c|}{0.4521}          & \multicolumn{1}{c|}{0.4288}          & \multicolumn{1}{c|}{0.4052}          & \multicolumn{1}{c|}{0.6567}          & 4.3580          & 0.5514                                                \\
                          & Mistral                 & 0.7631                                                    & \multicolumn{1}{c|}{0.9267}          & \multicolumn{1}{c|}{0.9040}          & \multicolumn{1}{c|}{0.9022}          & \multicolumn{1}{c|}{0.8602}          & \multicolumn{1}{c|}{\textbf{0.9388}} & \textbf{8.7706} & 90.8406                                               \\
                          & LLaMA-2                 & \textbf{0.7861}                                           & \multicolumn{1}{c|}{\textbf{0.9476}} & \multicolumn{1}{c|}{\textbf{0.9270}} & \multicolumn{1}{c|}{\textbf{0.9064}} & \multicolumn{1}{c|}{\textbf{0.8759}} & \multicolumn{1}{c|}{0.9274}          & 8.3282          & \textbf{92.8406}                                      \\ 
\bottomrule
\end{tabular}
\vspace{-0.15in}
\end{table*}

Table~\ref{Quality_Evaluation} present the results of various language models. The results show that:
1) LLMs Outperform Smaller Language Models. On the CWRU, Ottawa, and SCUT-FD datasets, large language models such as LLaMA-2 and Mistral consistently outperform GPT2-Large across all evaluation metrics. This demonstrates the stronger generalization capabilities of LLMs, especially in fault diagnosis tasks, where their performance is particularly prominent. 2) Pre-trained Models Excel in Cross-modal Alignment.
Mistral and LLaMA-2 achieve superior results on most evaluation metrics, including BLEU, ROUGE, CIDEr, and Match. This indicates that models pre-trained with general instructions are more adept at handling the spectrum-text alignment tasks in mechanical fault diagnosis. It further highlights the potential of instruction tuning in cross-modal tasks. 3) Performance on the Ottawa Dataset. FaultGPT exhibits lower performance on the Ottawa dataset compared to CWRU and SCUT-FD. This performance gap can be attributed to the smaller scale and potential quality limitations of the instruction-following data in the Ottawa dataset. This underscores the critical role that data scale and quality play in enhancing instruction-based fault diagnosis.

\subsection{Few-shot and Zero-shot Evaluation}
\begin{figure}[t]
    \centering
    \includegraphics[width=1\linewidth]{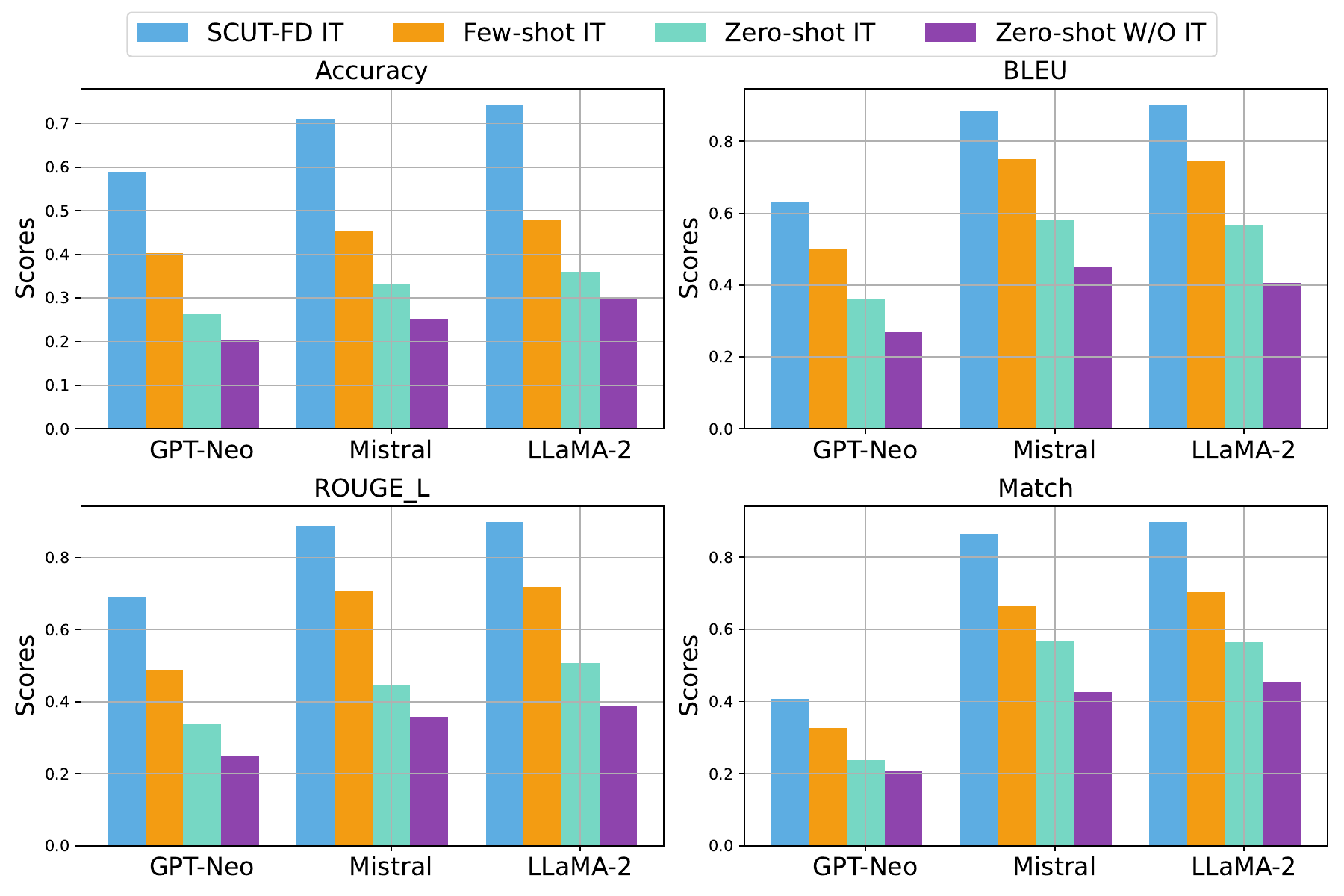}
    \caption{Few-shot and Zero-shot performance on SCUT-FD dataset. IT denotes instructing tuning.}
    \label{fig:zero_shot}
    \vspace{-0.15in}
\end{figure}

In Figure~\ref{fig:zero_shot}, we present the evaluation of the few-shot and zero-shot learning capabilities of various LLMs, which were trained on the CWRU dataset and then tested on the SCUT-FD (unseen dataset). The assessed models include GPT-Neo, Mistral, and LLaMA-2. Firstly, all selected LLMs undergo instruction tuning on the CWRU training set, followed by zero-shot testing on the SCUT-FD test set verified by human experts, denoted as \textit{Zero-shot Instruction Tuning (Zero-shot IT)}. We also measure the performance of each model in FDQA without prior FDQA-specific instruction tuning, denoted as \textit{Zero-shot without Instruction Tuning (Zero-shot W/O IT)}. 
Additionally, we assess the performance under a few-shot scenario, where each model is fine-tuned using a small subset (5-shot) of the SCUT-FD dataset before testing, denoted as \textit{Few-shot Instruction Tuning (Few-shot IT)}.
\textit{SCUT-FD IT} represents training on the SCUT-FD training set and then evaluating on the SCUT-FD test set.

Notably, although Few-shot IT and Zero-shot IT show a slight degradation compared to SCUT-FD IT, the results still indicate a strong generalization ability of the model to unseen datasets, especially in comparison to Zero-shot W/O IT. The involvement of FDQA instruction tuning enables the models to achieve superior zero-shot performance on the unseen SCUT-FD dataset, indicating the necessity of instruction tuning in enhancing the models' zero-shot ability in cross-dataset FDQA tasks.

\subsection{Ablation Study}
\subsubsection{Core Components in FaultGPT}
The MCID module we designed aims to extract fault semantics from time-frequency images and align them with mask images generated by Grad-CAM.
Therefore, inspired by anomaly detection techniques \cite{gu2024anomalygpt}, we adopt the Area Under the Receiver Operating Characteristic Curve (AUC) as an evaluation metric to assess the fault feature localization performance at the pixel level using Pixel-AUC. Meanwhile, token accuracy is defined as the proportion of correctly predicted tokens to the total number of target tokens. It serves as a metric to evaluate the alignment of the model's predictions with the expected outcomes.

The results of the ablation experiments are shown in Table~\ref{tab:ablation}.
Ablation studies demonstrate a consistent improvement in performance as components are incrementally introduced to the LLMs, starting with the MCID, followed by the prompt learner, and then the Adapter. For instance, on the SCUT-FD dataset, using only LLMs resulted in a Token Accuracy of 66.4\%, while the full model configuration achieved 85.6\%, signifying that the integration of all modules substantially increases the precision of fault diagnosis. Similarly, on the CWRU dataset, gains were notable, with Token Accuracy improving from 51.7\% with only LLMs to 86.3\% with the complete model.

\begin{table*}[t]
\centering
\caption{The impact of core components of FaultGPT on performance.}
\label{tab:ablation}
\setlength{\tabcolsep}{2mm}{
\begin{tabular}{cccccccccc}
\toprule
\multirow{2}{*}{LLMs}     & \multirow{2}{*}{MCID}     & \multirow{2}{*}{Prompt Learner} & \multirow{2}{*}{Adapter}  & \multicolumn{2}{c}{SCUT-FD}   & \multicolumn{2}{c}{CWRU}      & \multicolumn{2}{c}{Ottawa}    \\ \cline{5-10} 
                          &                           &                                 &                           & Pixel-AUC     & Token Acc     & Pixel-AUC     & Token Acc     & Pixel-AUC     & Token Acc     \\ \toprule
\checkmark &                           &                                 &                           & -             & 66.4          & -             & 51.7          & -             & 50.4          \\
\checkmark &                           & \checkmark       &                           & -             & 68.8          & -             & 55.1          & -             & 56.5          \\
\checkmark &                           &                                 & \checkmark & -             & 69.5          & -             & 56.1          & -             & 53.4          \\
\checkmark &                           & \checkmark       & \checkmark & -             & 70.7          & -             & 58.4          & -             & 59.2          \\ \midrule
\checkmark & \checkmark &                                 &                           & 88.7          & 72.2          & 83.6          & 66.3          & 80.2          & 66.5          \\
\checkmark & \checkmark & \checkmark       &                           & 89.2          & 77.3          & 86.7          & 75.3          & 84.5          & 75.4          \\
\checkmark & \checkmark &                                 & \checkmark & 88.7          & 79.5          & 83.6          & 76.4          & 80.2          & 73.7          \\ \midrule
\checkmark & \checkmark & \checkmark       & \checkmark & \textbf{92.8} & \textbf{85.6} & \textbf{90.6} & \textbf{86.3} & \textbf{85.1} & \textbf{77.8} \\ \bottomrule
\end{tabular}
}
\vspace{-0.15in}
\end{table*}

The combination of MCID and prompt learner alone achieved 89.2\% Pixel-AUC and 77.3\% Token Accuracy on the SCUT-FD dataset and 86.7\% Pixel-AUC and 75.3\% Token Accuracy on the CWRU dataset. The capability of MCID and prompt learner to guide LLMs in processing specific tasks significantly enhances the model's recognition and comprehension of features within fault images.

\subsubsection{Ablation Study of Loss Function}

\begin{table}[t]
\caption{Ablation study of loss function (SCUT-FD).} \label{ablation_loss}
\newcommand{\tabincell}[2]{\begin{tabular}{@{}#1@{}}#2\end{tabular}}
 \begin{center}
 \begin{threeparttable}
 \LARGE
    \resizebox{1.0\linewidth}{!}{
  \begin{tabular}{ccc|cc} \\
 	 \textit{CE} & \textit{Focal Loss} & \textit{Dice Loss} & \text{Pixel-AUC} ($\uparrow$) & \text{Token Acc} ($\uparrow$) \\
    \cmidrule{1-5}       
        \checkmark & ~ & ~ & -  & 66.4  \\ 
        \checkmark & \checkmark & ~ & 78.9  & 68.1  \\ 
        \checkmark & ~ & \checkmark & 83.4  & 69.8  \\ 
    \cmidrule{1-5}
        \checkmark & \checkmark & \checkmark & \textbf{88.7}  & \textbf{72.2} \\ 
	\end{tabular}
	}
	 \end{threeparttable}
	 \end{center}
  \vspace{-0.1in}
\end{table}   

To validate the effectiveness of different loss functions, we compared the individual and combined effects of cross entropy (CE) loss, focal loss, and dice oss in our experiments. Table~\ref{ablation_loss} presents the impact of these loss functions on fault diagnosis performance.
The experimental results show that incorporating dice loss yields significant performance improvements. With focal loss, Token Accuracy improved slightly, but the effect of dice loss was more pronounced, increasing Pixel-AUC from 78.9 to 83.4. This suggests that dice loss plays a crucial role in capturing fine-grained fault semantics. Moreover, the combination of CE Loss, focal loss, and dice loss achieved the best results, with Pixel-AUC increasing to 88.7 and Token Accuracy to 72.2, demonstrating the superior performance of multi-loss function fusion in fault diagnosis.

\subsubsection{Ablation Study on Instruction Tuning}
\begin{figure}[t]
    \centering
    \includegraphics[width=1\linewidth]{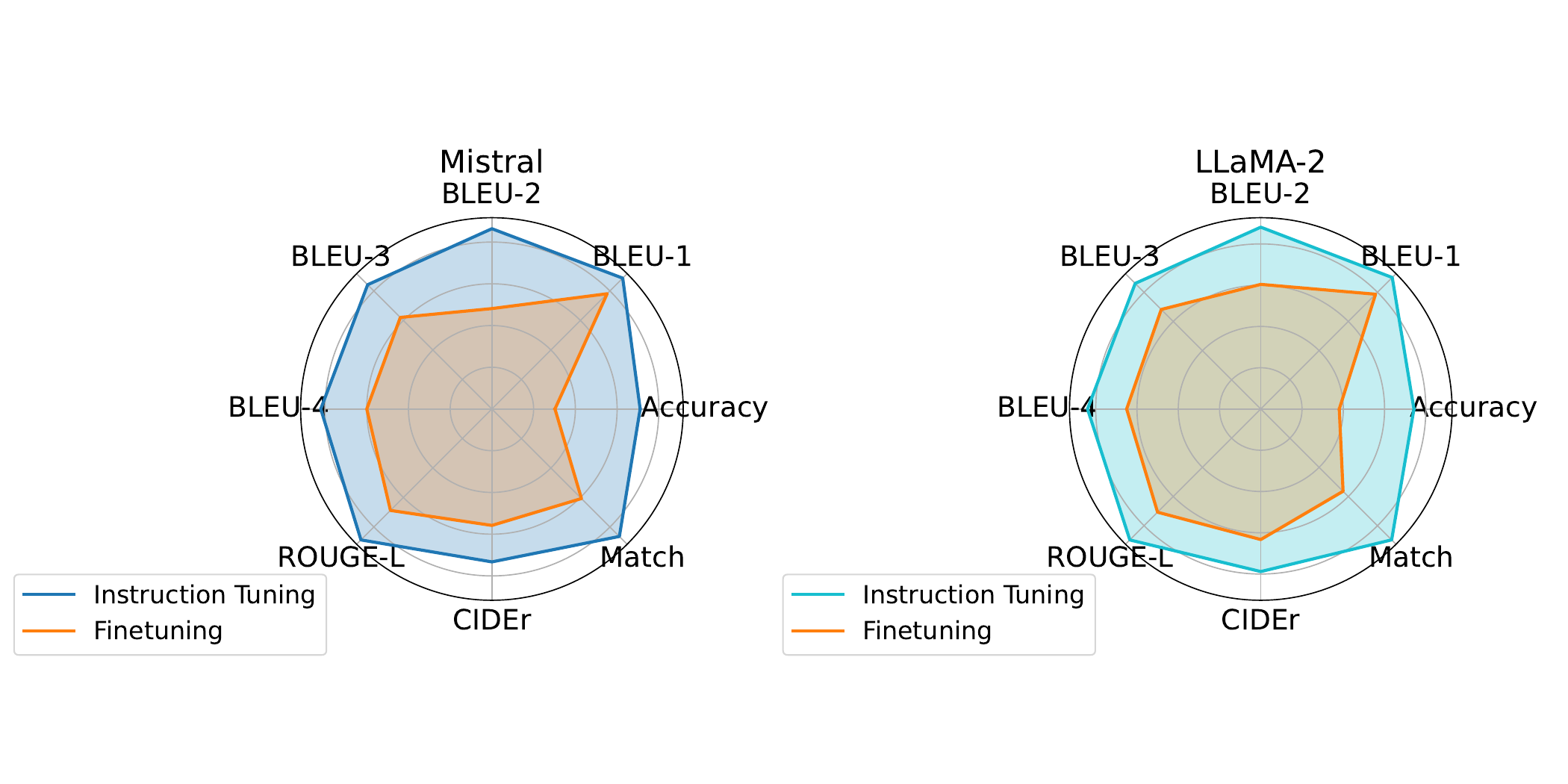}
    \caption{Ablation Study of Instruction Tuning on CWRU Dataset.}
    \label{fig:instruction_ft}
    \vspace{-0.15in}
\end{figure}

We conducted an ablation study to evaluate the impact of instruction tuning on the FDQA task. For the Mistral and LLaMA-2 models, we performed experiments where the models were trained directly on vibration time-frequency image-text pairs without instruction tuning. As shown in Fig.~\ref{fig:instruction_ft}, the results demonstrate that the performance of both models declines significantly across all metrics without instruction tuning, with a more pronounced drop observed in the Mistral model. This indicates that instruction tuning is more effective than direct fine-tuning in improving LLMs' generalization capabilities to new tasks or data, highlighting its crucial role in optimizing model performance.

\subsubsection{Ablation Study of Different Wavelet Bases}
The rolling bearing fault signals used in this paper are primarily characterized by impact components, and the fault information is mostly reflected in abrupt changes. Hence, the Morlet wavelet was chosen as the mother wavelet for time-frequency analysis in our experiments. 
We evaluated the performance of several commonly used wavelet bases, including Morlet, Haar, Mexican Hat, and Daubechies. As shown in Table~\ref{tab:CWT_ablation}, the performance of FaultGPT exhibited slight variations across different wavelet bases, indicating that the model is relatively insensitive to the choice of wavelet basis. Therefore, we selected the Morlet wavelet as the default configuration in our main experiments to ensure stable performance under different conditions. 

\begin{table}[t]
\caption{Ablation analyses of different wavelet bases. We report results on SCUT-FD.}
\label{tab:CWT_ablation}
\newcommand{\tabincell}[2]{\begin{tabular}{@{}#1@{}}#2\end{tabular}}
 \begin{center}
 \begin{threeparttable}
 \resizebox{1.0\linewidth}{!}{
  \begin{tabular}{lcccc}
  ~ & Morlet & Haar & Mexican Hat & Daubechies \\
  \cmidrule{1-5}
      Pix-AUC ($\uparrow$) & 88.7 & 88.0 & 88.2 & 87.5 \\
      Token Acc ($\uparrow$) & 72.2 & 71.1 & 71.6 & 70.8 \\
	\end{tabular}
 }
	 \end{threeparttable}
	 \end{center}
\end{table}

\subsection{User Interface Design}
\begin{figure}[t]
    \centering
    \includegraphics[width=0.9\linewidth]{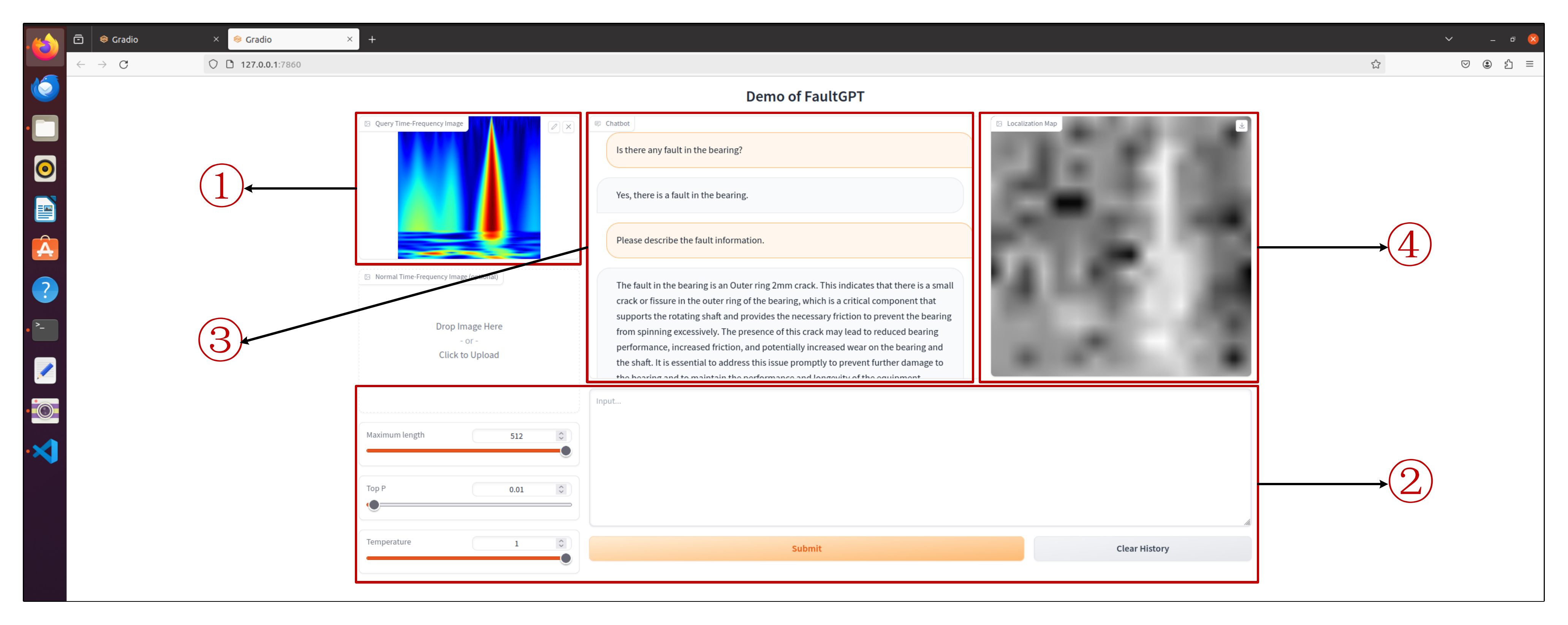}
    \caption{System demo showcasing an outer ring 2mm crack fault. The user interface is divided into four main sections: \ding{172} represents the time-frequency image input area, \ding{173} indicates the user instruction input and configuration selection area, \ding{174} is the diagnostic report generation area, and \ding{175} displays the feature maps extracted by the MCID.}
    \label{fig:abnormal demo}
    \vspace{-0.15in}
\end{figure}

\begin{figure}[t]
    \centering
    \includegraphics[width=0.9\linewidth]{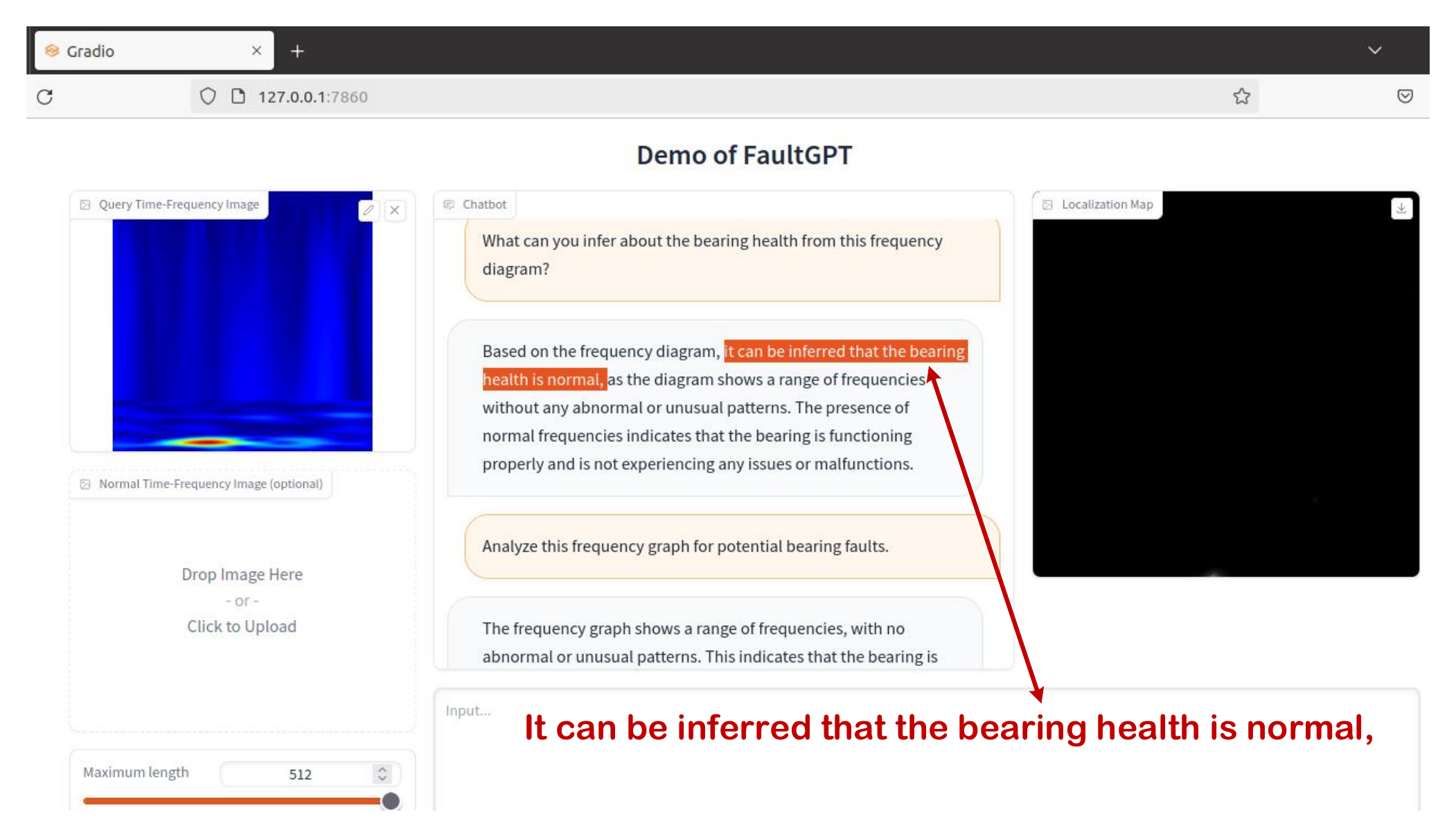}
    \caption{System demo showcasing a normal bearing. Since no faults are present, the MCID did not extract any fault-related semantics.}
    \label{fig:normal demo}
    \vspace{-0.15in}
\end{figure}

We develop a user-friendly interface to enable non-expert users to easily use the system for FD.
As illustrated in Fig.~\ref{fig:abnormal demo}-\ref{fig:normal demo}, our model is capable of detecting the presence of faults and, more intricately, estimating the severity level of these faults as well as determining their locations. Furthermore, users have the opportunity to engage in multiple rounds of dialogue concerning the content of the time-frequency images, facilitating a deeper interaction with the model's diagnostic process. \looseness=-1

We have developed a dedicated website to showcase our project, which provides detailed information and visual demonstrations. 

\section{Conclusion}
\label{sec:conclusion}
In this paper, we introduce fault diagnosis question answering, a novel research paradigm for industrial fault diagnosis. FaultGPT utilizes large vision language models  to achieve end-to-end, professional-grade generation of fault diagnosis reports, providing a more advanced and user-friendly alternative to traditional methods.
We constructed a multimodal instruction-tuning dataset and proposed an efficient and effective tuning method, enabling the model to accurately generate diagnostic reports based on vibration signals. Extensive experiments across three benchmark datasets highlighted FaultGPT’s strong performance in both generating fault diagnosis reports and conducting zero-shot testing.
Future research will explore the representation and report generation for compound fault diagnosis. 
Additionally, we aim to extend the application of FaultGPT to other industrial domains, such as predicting remaining useful life, further enhancing accuracy and adaptability across various industrial scenarios.

\bibliographystyle{IEEEtran}
\bibliography{IEEEabrv,references}
\appendix

\section*{Appendix A: Time-Frequency Imaging}
\label{cwt}
Time-frequency imaging is a technique that transforms raw sensor data into images by converting signal time series into the time-frequency domain \cite{10750065,shao2018highly}. This method is particularly useful in mechanical fault diagnosis as it provides a temporal understanding of the original data and allows signals to be associated with varying operational conditions. Time-frequency imaging can be achieved using various techniques, including Short-Time Fourier Transform (STFT) and Continuous Wavelet Transform (CWT).

STFT is implemented by applying a fixed window over the signal and performing a Fourier transform within the window, making it suitable for analyzing stationary signals. In contrast, CWT involves convolving the signal with a series of scaled and shifted mother wavelets. The mathematical expression for CWT is given by:
\begin{equation}
    \text{CWT}_x(a,b) = \frac{1}{\sqrt{|a|}} \int_{-\infty}^{\infty} x(t) \psi^* \left( \frac{t-b}{a} \right) dt,
\end{equation}
where \( x(t) \) is the signal to be analyzed, \( \psi(t) \) is the mother wavelet function, which must satisfy conditions such as finite energy and zero mean. The scale parameter \( a \) controls the compression or expansion of the wavelet, thus determining the frequency resolution. The shift parameter \( b \) represents the position of the wavelet in time, and \( ^* \) denotes the complex conjugate.

By adjusting the scale \( a \) and translation \( b \), CWT explores the signal’s frequency content across different scales and time positions, providing variable time-frequency resolution ideal for capturing transient features. These techniques allow one-dimensional time series data to be effectively transformed into time-frequency images, offering a richer perspective for analysis and fault diagnosis.

\section*{Appendix B: Gradient-weighted Class Activation Mapping}
\label{grad-cam}

To improve the extraction of fault features in time-frequency images, we utilize Grad-CAM to generate heatmaps, highlighting the areas most relevant to the fault diagnosis process. These heatmaps serve as a supervisory signal to enhance the accuracy of the MCID. By focusing on critical areas of the images, MCID is guided to extract detailed fault information, particularly when using focal loss and dice loss during training.

\begin{figure}[t]
    \centering
    \includegraphics[width=0.8\linewidth]{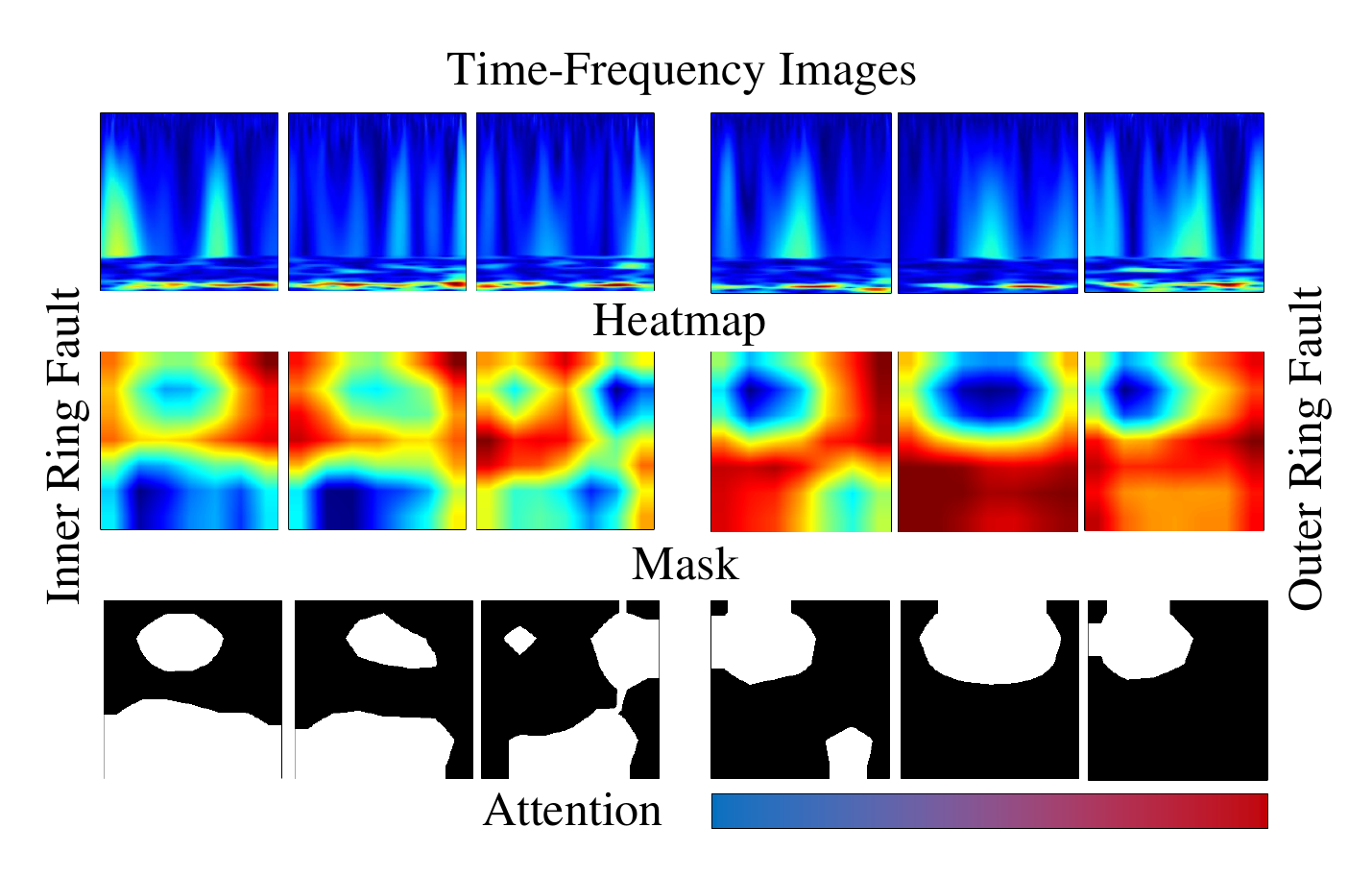}
    \caption{The heatmap of sampled vibration images, showcasing areas of attention for both inner and outer ring faults. The attention regions serve as critical guidance for extracting fault features in MCID.}
    \label{fig:heatmap}
    \vspace{-0.15in}
\end{figure}

Specifically, the heatmaps allow the model to concentrate on difficult-to-detect faults, while dice loss ensures precision in image segmentation. As shown in Fig.~\ref{fig:heatmap}, the heatmaps represent areas of high attention in the time-frequency images, aiding the learning process for both fault detection and segmentation.

The heatmaps displayed in Fig.~\ref{fig:heatmap} highlight the areas of focus for inner and outer ring faults. By incorporating these into the training process, we ensure that the model focuses on relevant regions during the fault classification and segmentation tasks.

\section*{Appendix C: Implementation Details}
\label{training_details}

We use the text encoder from CLIP \cite{radford2021learning} to process text prompts under normal and fault conditions, and ImageBind-Huge \cite{girdhar2023imagebind} as the image encoder to convert time-frequency representations of vibration signals into inferable feature vectors. The inference process is performed by the Vicuna-7B \cite{chiang2023vicuna} model by default, and in Section~\ref{llms}, we compare results from other LLMs. Text and image features are fused via a linear layer, providing rich contextual information for the large language model to generate accurate fault diagnosis results.

The model is initialized with pretrained parameters from PandaGPT \cite{su2023pandagpt}, significantly enhancing efficiency in fault diagnosis tasks. The image input resolution is fixed at 224×224 to ensure feature consistency. The training process spans 50 epochs with a learning rate set at 1e-3, and a batch size of 16. A linear warm-up and single-cycle cosine learning rate decay are used to prevent rapid learning in early training stages and mitigate overfitting in later stages.

During training, only the MCID module, Adapter module, and prompt learner are updated to ensure effective knowledge transfer from the pretrained model. The model is trained using not only the pretrained data from PandaGPT but also our custom time-frequency image-text dataset to adapt to more complex fault patterns. Training is conducted on two RTX-4090 GPUs (with 128GB physical memory) over 50 epochs, with the entire process completing within 10 hours.

\begin{figure}[ht]
    \centering
    \includegraphics[width=0.98\linewidth]{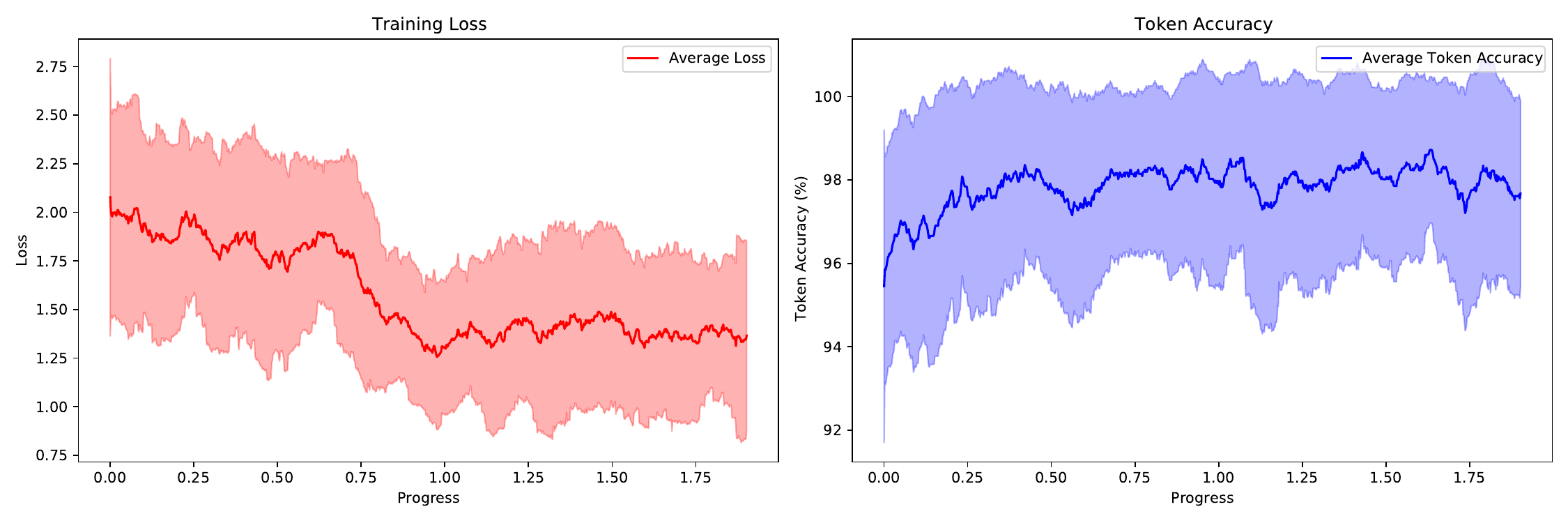}
    \caption{Mean Loss and Mean Token Accuracy for the training process.
    }
    \label{fig:training}
    \vspace{-0.15in}
\end{figure}

Fig.~\ref{fig:training} shows the loss and accuracy curves during training. In the inference phase, the model efficiently performs fault diagnosis by only requiring the time-frequency image of the sample.

\section*{Appendix D: Prompt learner.}
The specific design of prompt learner is shown in Table~\ref{tab:prompt learner}.

\begin{table}[th!]
\centering
\caption{Structure of the \texttt{PromptLearner} class.}
\begin{tabular}{lc}
\toprule
\textbf{Layer (Type)} & \textbf{Output Shape \& Details}        \\ \midrule
Input                 & (B, C, H, W)  \\
Conv2d                & (B, dim\_in*4, H, W), K=3, P=1  \\
Conv2d                & (B, dim\_in*16, H/2, W/2), K=3, P=1  \\
Conv2d                & (B, dim\_in*64, H/4, W/4), K=3, P=1  \\
Conv2d                & (B, dim\_in*256, H/8, W/8), K=3, P=1  \\
Conv2d                & (B, dim\_in*1024, H/16, W/16), K=3, P=1  \\
Conv2d                & (B, dim\_out, H/32, W/32), K=5, P=0  \\
Base Prompts          & (9, dim\_out) \\
\bottomrule
\end{tabular}
\label{tab:prompt learner}
\vspace{-0.15in}
\end{table}

\end{document}